\documentclass[prd,aps,nofootinbib,preprintnumbers]{revtex4}

\usepackage{graphicx}
\usepackage{amsmath}

\begin{document}

\preprint{ICRR-Report-627-2012-16,\ IPMU12-0179
}

\title{
$Q$ ball Decay Rates into Gravitinos and Quarks
}

\author{Masahiro Kawasaki$^{a,b}$ and Masaki Yamada$^{a}$}

\affiliation{
$^a$ Institute for Cosmic Ray Research, 
     University of Tokyo, Kashiwa, Chiba 277-8582, Japan\\
$^b$ Kavli Institute for the Physics and Mathematics of the Universe, 
     University of Tokyo, Kashiwa, Chiba 277-8582, Japan}

\date{\today}

\begin{abstract}
The Affleck-Dine mechanism, 
which is one of the most attractive candidates for the baryogenesis in supersymmetric theories, 
often predicts the existence of baryonic $Q$ balls in the early universe. 
In this scenario, there is a possibility to explain the observed baryon-to-dark matter ratio 
because $Q$ balls decay into supersymmetric particles as well as into quarks. 
If the gravitino mass is small compared to the typical interaction energy, 
the longitudinal component of the gravitino behaves like the massless goldstino. 
We numerically calculate the goldstino production rates from $Q$ balls 
in the leading semi-classical approximation without using large radius limit or effective coupling. 
We also calculate the quark production rates from $Q$ balls in the Yukawa theory with a massive fermion. 
In deriving the decay rate we also take into account the scalar field configuration of the $Q$ ball. 
These results are applied to a realistic model in the gauge-mediated supersymmetry breaking 
and yield the branching ratio of the $Q$ ball decay into the gravitino. 
We obtain the branching ratio much smaller than the one estimated in the previous analysis. 
\end{abstract}

\maketitle

\section{\label{sec1}Introduction}
In cosmology, the origin of dark matter and the baryon asymmetry is very important problem 
which can not be explained by the Standard Model of elementary particle physics. 
Supersymmetric (SUSY) extensions of the Standard Model can explain this problem. 
In minimal supersymmetric standard model (MSSM) the lightest 
SUSY particle (LSP) is stable and is natural candidate for the dark matter. 
Besides, the baryon asymmetry can be produced by the Affleck-Dine mechanism~\cite{AD,DRT}. 
The Affleck-Dine mechanism produces a scalar field condensate with baryon number. 
In many models such as the gauge-mediated SUSY breaking (GMSB) models, 
this condensate feels spatial instabilities and 
fragments into nontopological solitons, $Q$ balls~\cite{KuSh,EnMc,KK1,KK2,KK3}. 

In the GMSB, the LSP is gravitino. 
If the charge of the $Q$ ball is large enough, 
the $Q$ ball is stable. Since the stable $Q$ ball has astrophysical problems~\cite{KLS1, KLS2}, 
we do not consider this case. 
On the other hand, if the charge of the $Q$ ball is small enough, 
the $Q$ ball decay into hadrons and gravitinos. 
If $Q$ balls decay only via squark $\to$ quark $+$ gravitino~\cite{ShKu,DoMc}, 
the produced gravitino number density is equal to 
the produced quark number density by R-parity conservation. 
Then the gravitino mass should be $\simeq 1.7$GeV 
to explain the observed baryon-to-dark matter ratio $\simeq 1/5$. 
However, it is pointed out that the process as squark $+$ squark $\to$ quark $+$ quark occurs via heavy gluino exchange 
and is main decay mode if decay into a squark is kinematically forbidden~\cite{KK2011}. 
In this case, it becomes important to calculate the branching ratios into the gravitino and quarks 
to estimate the baryon-to-dark matter ratio. 

In this article, therefore, 
we derive the production rates of gravitinos and quarks from the $Q$ ball 
in the GMSB model (the most interesting case) and in the gravity-mediated SUSY breaking model. 
The $Q$ ball decay was first studied by Cohen \textit{et al.}~\cite{evap}, who 
considered the Yukawa theory and treated the scalar field as the classical 
$Q$ ball background field in the leading semi-classical approximation 
and calculated the fermion production rate by the Bogoliubov transformation 
between creation and annihilation operators at $t \to \pm \infty$~\cite{evap}. 
In the case of gravitino, 
if the gravitino mass is small compared to the typical interaction energy, 
the longitudinal component of the gravitino behaves like the massless goldstino 
and has the derivative interactions with chiral multiplets. 
Therefore we consider fermion fields (one of which is a goldstino) with derivative coupling to a scalar field. 
Once we treat the scalar field as the classical background field, 
the interaction terms become at most second order for the fermion fields and 
this system can be solved numerically in the same way as Cohen \textit{et al.} did. 
The gravitino production rate from $Q$ balls was first calculated in Ref.~\cite{KK2011}, 
where the effective coupling estimated from the decay rate of 
squark $\to$ quark $+$ gravitino and large radius limit are used. 
We derive the gravitino production rate without relying on the effective coupling and large radius limit. 
We also consider two fermion fields one of which has a mass term in the Yukawa theory. 
This system is almost equivalent to the one considered in Refs.~\cite{KLS1, HNO}. 
The interaction of $Q$ balls with ordinary matter was studied 
in Ref.~\cite{KLS1}, which showed that 
quarks are reflected as antiquarks with a probability $\sim 1$ 
but did not refer to the $Q$ ball decay. 
In Ref.~\cite{HNO}, the authors estimated the quark production rate 
using the effective interaction after integrating out the heavy particle. 
However, in this paper we calculate the rate of the $Q$ ball decay into quarks 
without integrating out the heavy particle. 
Furthermore, in deriving the decay rate 
we also take the scalar field configuration of the $Q$ ball into account. 
In most of the previous studies, a step function is used as the scalar field configuration. 
We carefully examine the differences 
between the realistic and step-function configurations. 

This paper is organized as follows. In Sec.~\ref{sec2}, we briefly review the property of $Q$ balls 
in the GMSB and the gravity mediated SUSY breaking models. 
In Sec.~\ref{sec3}, we review the method to calculate the decay rate of $Q$ balls and 
discuss the differences among the decay rates for some $Q$ ball configurations in the Yukawa theory. 
We apply the method to calculate the goldstino production rate from the $Q$ balls 
in the GMSB and in the gravity mediated SUSY breaking models in Sec.~\ref{sec4}. 
In Sec.~\ref{sec5}, we also calculate the $Q$ ball decay rate in a theory with a massive fermion. 
Then, we apply our results to gravitino and quark production from $Q$ balls in Sec.~\ref{sec6}. 
Sec.~\ref{sec7} is devoted to the conclusion.

\section{\label{sec2}Q ball solutions}
In this section, we consider a complex scalar field theory with a global $U(1)$ symmetry. 
The Lagrangian density is written as 
\begin{equation}
\mathcal{L} = - \partial_\mu \phi^* \partial^\mu \phi - V(\phi).
\end{equation}
The conserved $U(1)$ charge and the energy are given by 
\begin{eqnarray}
&&Q = i \int \left( \phi^* \partial_0 \phi - \phi \partial_0 \phi^* \right) d^3x, \\
&&E = \int \left( \partial_0 \phi^* \partial_0 \phi + \partial_i \phi^* \partial_i \phi + V \left( \phi \right) \right) d^3x, 
\end{eqnarray}
respectively. 
The scalar field configuration which minimizes the energy at a fixed charge $Q$ is obtained by minimizing 
\begin{equation}
E + \omega_0 \left( Q - i \int \left( \phi^* \partial_0 \phi - \phi \partial_0 \phi^* \right) d^3x \right), 
\end{equation}
where $\omega_0$ is a Lagrange multiplier. 
We can immediately determine the time dependence of the scalar field configuration 
and obtain $\phi({\bf r},t)=\phi({\bf r}) e^{-i \omega_0 t}$. 
Then, taking a spherically symmetric ansatz $\phi( {\bf r})=\phi(r )$, 
we get the radial part of the configuration by solving the following equation: 
\begin{equation}
\frac{\partial^2}{\partial r^2} \phi + \frac{2}{r} \frac{\partial}{\partial r} \phi + \omega_0^2 \phi - 
\frac{1}{2} \frac{\partial}{\partial \phi} V ( \phi ) = 0, 
\label{phi solve}
\end{equation}
with the boundary condition $\phi'(0 )=0$. 
Here, if $\omega_0^2 < \mathrm{min}_\phi \left( \frac{V}{\vert \phi \vert^2} \right) < V''(0)/2$, 
we get the spatially localized configuration, called $Q$ ball~\cite{Coleman}. 

In MSSM, there are many flat directions in the scalar potential. 
The flat directions are combinations of squarks, sleptons and Higgs, 
but they are lifted by the SUSY breaking effect. 
In the early universe, the Affleck-Dine mechanism produces 
a scalar field condensate with baryon number using one of these flat directions. 
If a $Q$ ball solution exists for this flat direction, 
the scalar field condensate fragments into $Q$ balls~\cite{Qsusy}. 
In the following subsections, we review the properties of the $Q$ ball in 
the GMSB 
and the gravity-mediated supersymmetry breaking models. 

\subsection{\label{sec2-1}Properties of the $Q$ ball in gauge mediation}
\begin{figure}[htbp]
\begin{tabular}{c}
 \includegraphics[width=100mm]{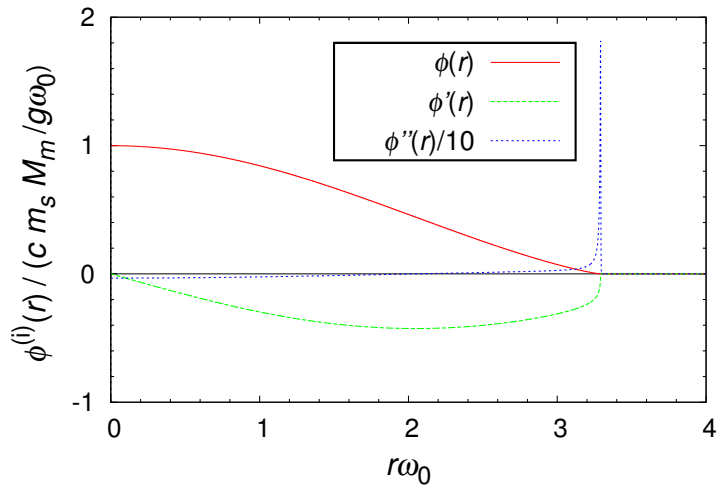}
\end{tabular}
\vspace{-0.3cm}
\caption{Numerical solutions of Eq.~(\ref{phi solve}), $\phi( r)$, $\phi'( r)$ and $\phi''( r)$, 
in the GMSB with $m_s / \omega_0=10^3$. 
The vertical axis is normalized by $\left( c m_s M_m / g\omega_0 \right) $ for $\phi( r)$ and $\phi'( r)$, 
and $\left( c m_s M_m /g\omega_0 \times 1/10 \right) $ for $\phi''( r)$. 
The horizontal axis denotes the radius normalized by $1/\omega_0$. 
}
\label{GMSBphi}
\end{figure}

In the GMSB model, the gauge fields acquire large masses for $g \phi \gg M_m$, 
where $M_m$ is the messenger scale, and $g$ generically stands for the standard model gauge coupling. 
Then the transmission of SUSY breaking effect is suppressed 
and the flat directions flatten out for $g \phi \gg M_m$ as~\cite{log2} 
\begin{equation}
V= \frac{2 m^2_s M_m^2}{g^2} \left[ \log \left( 1+ \frac{g \vert \phi \vert }{M_m} \right) \right]^2 , 
\label{GMSB V}
\end{equation}
where $m_s$ is the soft mass scale. 
In the GMSB gravitino mass $m_{3/2}$ is much smaller than $m_s$. 
In this potential, there always exists a $Q$ ball solution. 
The mass of $Q$ ball $M_Q$, the typical size of the $Q$ ball $R$, $\omega_0$, 
and the field value at the center of the $Q$ ball $\phi_0 \equiv \phi (r=0)$ are written as 
\begin{align}
&M_Q \simeq \frac{4 \sqrt{2 \pi c}}{3} \left( \frac{m_s M_m}{g} \right) ^{1/2} Q^{3/4},  \\ 
&R \simeq \sqrt{ \frac{\pi}{2 c}} \left( \frac{g }{m_s M_m} \right)^{1/2}  Q^{1/4}
\left( \simeq \frac{ \pi }{ \omega_0} \right),  \label{GMSBR} \\
&\omega_0 \simeq \sqrt{2 \pi c} \left( \frac{m_s M_m}{g} \right)^{1/2} Q^{-1/4},  \\ 
&\phi_0 \simeq \sqrt{\frac{c}{2 \pi}} \left(  \frac{m_s M_m}{g} \right) ^{1/2} Q^{1/4} 
\left( \simeq c \frac{m_s M_m}{g \omega_0} \right) \label{GMSB property},  
\end{align}
where the parameter $c$ is fitted as~\cite{HNO} 
\begin{equation}
c \simeq 4.8 \log (m_s / \omega_0) + 7.4. 
\label{HNO const}
\end{equation}
The scalar filed configuration is well approximated by 
$\phi=\phi_0 \frac{\sin \left(  \omega_0 r \right) }{\omega_0 r }$ for $r< R$. 
However, this configuration is not smooth at $r=R$ [see Eq.~(\ref{GMSBR})]. 
This cause some difficulty in deriving decay rate of the $Q$ ball into goldstinos. 
Thus in order to obtain the smooth configuration we solve Eqs.~(\ref{phi solve})~and~(\ref{GMSB V}) numerically. 
We show the numerical solution in Fig.~\ref{GMSBphi}. 
The second derivative drastically changes around $r\omega_0 \sim \pi$, but it is finite and nonsingular, as expected. 

\subsection{\label{sec2-2}Properties of the $Q$ ball in gravity mediation}
In the gravity-mediated SUSY breaking model, the flat directions are lifted by the SUSY breaking effect as 
\begin{equation}
V = m^2_s \vert \phi \vert^2 \left( 1+ K \log \frac{ \vert \phi \vert^2}{M^2_G} \right) , 
\end{equation}
where $m_s$ is the soft mass scale and $M_G$ is the reduced Planck mass. 
In the gravity-mediated SUSY breaking gravitino mass $m_{3/2}$ is the same order of $m_s$. 
The second term in the parenthesis comes from radiative correction, and typically $ \vert K \vert  \sim$0.01-0.1. 
It is known that $K<0$ for the flat directions of the first and second generation squarks, 
and then there exists a $Q$ ball solution~\cite{EnMc}. 
The scalar field configuration is well approximated by $\phi(r ) \simeq \phi_0 \exp \left( - \frac{r^2}{2 R^2} \right)$, 
and $M_Q$, $R$, $\omega_0$ and $\phi_0$ are written as 
\begin{align}
&M_Q \simeq m_s Q, \\
&R \simeq \frac{1}{ \vert K \vert^{1/2} m_s}, \label{gravproperty} \\
& \omega_0 \simeq m_s, \\
&\phi_0 \simeq (2 \pi^{3/2} )^{-1/2} \vert K \vert^{3/4} m_s Q^{1/2},  
\end{align}
%

\section{\label{sec3}Q ball decay in the Yukawa theory}
In this section, we consider the Yukawa theory in the $Q$ ball background field. 
In this system, Cohen \textit{et al.} proved that 
the $Q$ ball decay into fermions in the leading semi-classical approximation. 
Here, we first review the method presented by them and then apply to some cases without using large radius limit. 

The Lagrangian density is written as 
\begin{equation}
\mathcal{L} = \chi^\dagger i \Bar{ \sigma }^\mu \partial_\mu \chi 
+ \eta^\dagger i \Bar{ \sigma }^\mu \partial_\mu \eta
- \left( g \phi^{*} \chi  \eta + h.c. \right), 
\label{Coleman lagrangian}
\end{equation}
where $\chi$ and $\eta$ are two-component Weyl spinors, and 
$\bar{\sigma}^\mu=( \bf{1}, -\sigma^i)$, where $\sigma^i$'s are the Pauli matrices. 
We assign the global $U(1)$ charge for $\phi$, $\chi$ and $\eta$ as 1, 1 and 0, respectively, 
and treat the scalar field as the $Q$ ball classical background field 
$\phi= \phi( r) e^{-i\omega_0 t}$ in the leading semi-classical approximation. 
The discussion below is also correct for more general interactions like the one used in the following sections. 

When we treat the scalar field as the classical background field, 
this system is invariant under simultaneous time translations and $U(1)$ rotations as 
\begin{equation}
\begin{cases}
t \to t +\delta t, \\
\chi \to \chi e^{-i\omega_0 \delta t}.
\end{cases}
\end{equation}
The associated conserved current is 
\begin{equation}
\begin{split}
 j^\mu &= \chi^\dagger i \Bar{\sigma}^\mu \left( \partial_0 + i \omega_0 \right)  \chi +
 \eta^\dagger i \Bar{\sigma}^\mu \partial_0 \eta 
- \delta^\mu_0 \mathcal{L}, \\
&=T^\mu_0 -\omega_0 j^\mu, 
\end{split}
\end{equation}
where $T^\mu_\nu$ is the energy momentum tensor for the fermions and $j^\mu$ is the $U(1)$ current. 
Then, the conserved charge is not the energy $E_\mathrm{f}$ itself but $E_\mathrm{f}-\omega_0 Q_\mathrm{f}$, 
where $Q_\mathrm{f}$ is the $U(1)$ charge for the fermions. 
Using this conservation law, we can show that the fermion fields are created 
from the $Q$ ball surface in the leading semi-classical approximation. 
We have treated the scalar field as the $Q$ ball classical background field 
and the Lagrangian density is at most second order for the fermion fields, 
so we can easily solve this quantized fermionic system. 
Since the conserved charge is not $E_\mathrm{f}$ but $E_\mathrm{f}-\omega_0 Q_\mathrm{f}$, 
positive and negative modes may mix with each other in this system. 
Thus, even if an incoming wave is a positive mode (annihilation operator) at $t=- \infty$, 
the outgoing wave can be a negative mode (creation operator) at $t= \infty$. 
We can regard this as the Bogoliubov transformation between annihilation and creation operators at $t  \to \pm \infty$, 
and then we can calculate the non-zero number density of outgoing waves. 

We need the free field mode expansions of the fermions 
to consider the scattering problem of the fermion fields in the $Q$ ball background field. 
Far away from the $Q$ ball surface, the background field is $\phi=0$, 
and then the equation of motion for the mode of $\chi \propto e^{-i k_+ t}$ is 
\begin{equation}
\left( k_+ - i {\bf \sigma} \cdot {\bf \nabla} \right) \chi =0. 
\end{equation}
In this article, we treat $\chi_\alpha$, $\chi^{\dagger \dot{\alpha}}$, 
$\eta_\alpha$ and $\eta^{\dagger \dot{\alpha}}$ as column vectors and neglect the indices. 
Due to the rotational invariance of this system, we can expand the solutions by the following Pauli spinors: 
\begin{equation}
\begin{cases}
\Phi \left( j, m, l = j+1/2 \right)  =
\begin{pmatrix}
\frac{\sqrt{j+1-m}}{\sqrt{2(j+1)}} Y^{m-1/2}_l \\[3mm]
-\frac{\sqrt{j+1+m}}{\sqrt{2(j+1)}} Y^{m+1/2}_l 
\end{pmatrix}, \\[8mm] 
\Phi \left( j, m, l' = j-1/2 \right)  =
\begin{pmatrix}
\frac{\sqrt{j+m}}{\sqrt{2j}} Y^{m-1/2}_{l'} \\[3mm]
\frac{\sqrt{j-m}}{\sqrt{2j}} Y^{m+1/2}_{l'} 
\end{pmatrix}, 
\end{cases}
\end{equation}
where $Y^m_l$ are the spherical harmonics. 
Defining $u^{(i)}$ (i=1, 2) as 
\begin{equation}
u^{(i)}(k,j,m;{\bf r} ) = \frac{k}{\sqrt{\pi}} [ h^{(i)}_{l'} (kr) \Phi(j,m,l') + h^{(i)}_l (kr) \Phi(j,m,l) ], 
\label{u hankel} 
\end{equation}
where $h^{(1)}_l$ and $h^{(2)}_l$ are the spherical Hankel functions of the first and second kinds, respectively, 
we have $\left(  k+ i {\bf \sigma} \cdot \nabla \right)  u^{(i)}(k) = 0$. 
So we obtain the solutions outside the $Q$ ball as 
\begin{multline}
\chi = \sum_{j,m} \int^{\infty}_0 dk_+
 \{ 
a_{\mathrm{in}}(k_+,j,m)e^{-ik_+t}u^{(1)}(-k_+,j,m;{\bf r})  \\
+ a_{\mathrm{out}} (k_+,j,m)e^{-ik_+t} u^{(2)} (-k_+,j,m;{\bf r})  
+ \mathrm{(terms\ of\ antiparticles)} 
\}, 
\end{multline}
where $a_{\mathrm{in}}$ and $a_{\mathrm{out}}$ are expansion coefficients. 
We expand $\eta$ in the same way as $\chi$ but denote expansion coefficients as $c$ instead of  $a$. 
For later use, we write the expansion of $\eta^\dagger \equiv \eta^{\dagger \dot{\alpha}}$ as 
\begin{multline}
\eta^{\dagger}=i \sigma_2 \left(  \eta_\alpha \right) ^* = \sum_{j,m} \int^{\infty}_0 dk_+
\{ 
c_{\mathrm{in}}^\dagger(k_+,j,-m)e^{ik_+t}(-1)^{m_-}u^{(2)}(-k_+,j,m;{\bf r})  \\
+ c_{\mathrm{out}}^\dagger (k_+,j,-m)e^{ik_+t} (-1)^{m_-} u^{(1)} (-k_+,j,m;{\bf r})  
+ \mathrm{(terms\ of\ antiparticles)} 
\}, 
\label{eta expansion}
\end{multline}
where we have used $i\sigma_2 u^{(1,2)} ( k,j,m; {\bf r})^* = (-1)^{m_+} u^{(2,1)} ( k,j,-m; {\bf r})$, 
where $m_\pm \equiv m \pm 1/2$. 

When we quantize the fields, $a_{\mathrm{in}}$ and $a_{\mathrm{out}}$ become the annihilation operators 
for incoming and outgoing waves, respectively. 
The operators obey the Heisenberg equations of motion and are scattered by the $Q$ ball at the origin. 
We shall consider this scattering problem. 

First, let us consider the case without $Q$ balls. 
Since the solution has to be nonsingular at the origin, 
the radial part of the solution is written by the spherical Bessel functions, $j_l(kr)$, as $\chi \propto u(-k,j,m;{\bf r} )$, 
where $u(k,j,m;{\bf r} )$ is defined by substituting the spherical Bessel functions 
for the spherical Hankel functions in the definition of $u^{(i)}$. 
Using this, along with $j_l=(h^{(1)}_l +h^{(2)}_l)/2$, we obtain $a_{\mathrm{out}}=a_{\mathrm{in}}$. 
This means that the incoming wave reflects off the origin and becomes the outgoing wave. 
Of course, the number density of the outgoing wave is 
$\langle 0_{\mathrm{in}} \vert  a^\dagger_{\mathrm{out}}a_{\mathrm{out}}  \vert 0_{\mathrm{in}} \rangle = 0$ 
when the vacuum, $ \vert  0_{\mathrm{in}} \rangle$, is defined as the state of no incoming wave, 
i.e. $a_{\mathrm{in}}  \vert  0_{\mathrm{in}} \rangle=0$. 

Next we consider the case with the $Q$ ball background field whose center is located at the origin of the coordinate. 
The Heisenberg equations of motion are 
\begin{equation}
\begin{cases}
\displaystyle{
i \Bar{\sigma}^{\mu}\partial_{\mu} \chi -g \phi \eta^\dagger  =0}, \\[4mm]
\displaystyle{
i {\sigma}^\mu \partial_\mu \eta^\dagger -g \phi^* \chi =0}. 
\end{cases}
\label{EOM Yukawa}
\end{equation}
Since this is the combination of the linear differential equations, 
we can solve these equations with appropriate boundary condition, 
and then outgoing waves,  $a_{\mathrm{out}}$ and $c_{\mathrm{out}}$, 
can be written by incoming waves, $a_{\mathrm{in}}$ and $c_{\mathrm{in}}$, like the previous example. 
However, because the background field depends on the time as $\phi \propto e^{-i \omega_0 t}$, 
the conserved quantity is not $E_\mathrm{f}$ but $E_\mathrm{f}-\omega_0 Q_\mathrm{f}$. 
Thus, the modes which mix with each other are 
\begin{equation}
\begin{cases}
\chi \propto e^{-i\omega t} = e^{-ik_+ t }, \\
\eta^\dagger \propto e^{-i(\omega - \omega_0) t }  = e^{ik_- t }. 
\end{cases}
\label{mode dif}
\end{equation}
We are interested in situations where positive and negative modes mix with each other, 
because we derive the fermion production rate through the Bogoliubov transformation 
between creation and annihilation operators. 
Thus, we restrict our attention to $0 < \omega < \omega_0$. 

Outside the $Q$ ball, the fields satisfy the free equations of motion. 
The angular momentum conservation implies that the terms which mix with each other can be written by 
\begin{equation}
\begin{cases}
a_{\mathrm{in}}(k_+,j,m) e^{-ik_+t} u^{(1)}(-k_+,j,m;{\bf r} ) + a_{\mathrm{out}} (k_+,j,m) e^{-ik_+t} u^{(2)}(-k_+,j,m;{\bf r} ), \\
c^\dagger_{\mathrm{in}} (k_-,j,-m) e^{ik_-t} (-1)^{m_-} u^{(2)}(-k_-,j,m,;{\bf r}) + 
c^\dagger_{\mathrm{out}} (k_-,j,-m) e^{ik_-t} (-1)^{m_-} u^{(1)}(-k_-,j,m,;{\bf r}), 
\end{cases}
\label{mix mode}
\end{equation}
where $a_{\mathrm{out}}$ and $c^\dagger_{\mathrm{out}}$ 
can be written by $a_{\mathrm{in}}$ and  $c^\dagger_{\mathrm{in}}$ by matching the interior and exterior solutions. 
Using the superposition principle, we can write 
\begin{equation}
\begin{pmatrix}
a_{\mathrm{out}}(k_+,j,m) \\
(-1)^{m_-} c^\dagger_{\mathrm{out}}(k_-,j,-m)  \\
\end{pmatrix}
=
\begin{pmatrix}
R_\chi(k_+,j)	&T_\chi(k_+,j)	\\
T_\eta(k_-,j)	&R_\eta (k_-,j)	
\end{pmatrix}
\begin{pmatrix}
a_{\mathrm{in}}(k_+,j,m) \\
 (-1)^{m_-}  c^\dagger_{\mathrm{in}}(k_-,j,-m) \\
\end{pmatrix}, 
\label{bogo}
\end{equation}
where $R_i$ and $T_i$ are coefficients fixed later by matching the interior and exterior solutions 
and do not depend on $m$ due to rotational invariance. 
We can regard this as the Bogoliubov transformation between $t \to \pm \infty$. 
The anticommutation relations 
$\{a_{\mathrm{in}}, a^\dagger_{\mathrm{in}}\}=\{a_{\mathrm{out}}, a^\dagger_{\mathrm{out}}\} = 
\{c_{\mathrm{in}},c^\dagger_{\mathrm{in}}\}=\{c_{\mathrm{out}},c^\dagger_{\mathrm{out}}\}$ 
imply that this translation matrix is a unitary matrix. 
Especially, we have 
\begin{equation}
\begin{cases}
\vert T_\chi(k_+,j) \vert^2 = \vert T_\eta(k_-,j) \vert^2,  \\
\vert R_i (k_+,j) \vert^2 + \vert T_i(k_-,j) \vert^2=1, \qquad (i=\chi , \eta).
\end{cases}
\label{dif unitarity}
\end{equation}

We define the vacuum, $\vert 0_{\mathrm{in}} \rangle$, 
as $a_{\mathrm{in}} \vert 0_{\mathrm{in}} \rangle  =c_{\mathrm{in}} \vert 0_{\mathrm{in}} \rangle  =0$ at $r\to \infty$, 
where the incoming waves are not affected by the $Q$ ball. 
The incoming waves are scattered by the $Q$ ball and 
this scattering process is described by the Heisenberg equation of motion. 
Then, we get Eq.~(\ref{bogo}) and the number density of the outgoing $\chi$ waves is 
\begin{equation}
\begin{split}
\langle 0_{\mathrm{in}}  \vert a_{\mathrm{out}}^\dagger(k_+,j,m) a_{\mathrm{out}}(k'_+,j',m') \vert 0_{\mathrm{in}} \rangle 
&=(-1)^{m_- +m'_-} T_\chi^*(k_+,j)T_\chi(k'_+,j')  \langle 0_{\mathrm{in}}  
\vert  c_{\mathrm{in}}(k_-,j,-m) c_{\mathrm{in}}^\dagger(k'_-,j',-m') \vert 0_{\mathrm{in}} \rangle, \\ 
&= \vert T_\chi(k_+) \vert^2\delta(k_+ -k'_+) \delta_{j,j'} \delta_{m,m'}. 
\end{split}
\end{equation}
This proves that the outgoing $\chi$ waves are created by the presence of the $Q$ ball background field. 
Similarly, the outgoing $\eta$ waves are also created. 
The above unitarity condition, 
$\vert T_\chi(k_+,j) \vert^2= \vert T_\eta(k_-,j) \vert^2$, of Eq.~(\ref{dif unitarity}) 
implies that the process can be represented by 
$\phi_{BG}(E=\omega_0) \to \chi(E=k_+) + \eta(E=k_- \equiv \omega_0 - k_+)$, 
where $\phi_{BG}$ is the $Q$ ball background field, and $E$ is the energy of each field. 
The second condition, 
$\vert R_\chi (k_+,j) \vert^2 + \vert T_\chi(k_-,j) \vert^2=1$, of Eq.~(\ref{dif unitarity}) 
implies that the production rate is bounded above due to the Pauli exclusion principle. 
Summing over the states and using $\delta(0)=T/ 2\pi$, 
we obtain the following production rates of the fields from the $Q$ ball: 
\begin{equation}
\frac{d}{dt}N_i = \sum_{j=1/2} \int^{\omega_0}_0 \frac{dk}{2\pi} (2j+1) \vert T_i (k,j) \vert^2, 
\qquad (i=\chi, \eta). 
\label{evaporationrate}
\end{equation}

Finally, we need to determine the coefficients $T_i$ by matching the interior and exterior solutions. 
Inside the $Q$ ball, the Heisenberg equations of motion are given by Eq.~(\ref{EOMdiff}). 
With use of Eq.~(\ref{mode dif}), the equations are rewritten as 
\begin{equation}
\begin{cases}
\displaystyle{
\left( k_+ -i {\bf \sigma \cdot \nabla} \right) \chi 
-g \phi(r ) \eta^\dagger =0}, \\[4mm]
\displaystyle{ 
\left( -k_- +i {\bf \sigma \cdot \nabla} \right) \eta^\dagger 
-g \phi(r ) \chi =0}, 
\end{cases}
\label{EOM Yukawa 2}
\end{equation}
The conservation of the angular momentum implies that the solutions can be expanded as~\cite{MV} 
\begin{equation} 
\begin{cases}
\chi = f_\chi \left( r \right) \Phi \left( j,m,l' \right) + i g_\chi \left( r \right) \Phi \left( j,m,l \right), \\
\eta^\dagger = f_\eta \left( r \right) \Phi \left( j,m,l' \right) + i g_\eta \left( r \right) \Phi \left( j,m,l \right). 
\end{cases}
\end{equation}
Then, using 
\begin{equation}
{\bf \sigma \cdot \nabla} \Phi \left( j,m, j \pm 1/2 \right)
=  \Phi \left( j,m, j \mp 1/2 \right) \times \left(  \frac{\partial}{\partial r} + \frac{1 \pm \left( j+1/2 \right) }{r} \right), 
\label{formula1}
\end{equation}
we obtain the four first order differential equations. 
Since the solutions have to be nonsingular at $r=0$, the boundary conditions are 
$f'_i \left( r=0 \right) =g'_i \left( r=0 \right) =0$ for $i=\chi$ and $\eta$. 
We can get two independent solutions numerically for the given scalar field configuration $\phi( r)$. 
Then, matching the interior and exterior solutions and using Eqs.~(\ref{mix mode})~and~(\ref{bogo}), 
we obtain the coefficients $T_i$. 
The matching condition is simply setting the solutions equal to each other at sufficiently large $r$ 
where $\phi( r) \simeq 0$ is satisfied. 

\subsection{\label{sec3-1}Q ball decay rate for $R\omega_0 \to \infty$ in the Yukawa theory}
Using the above technique and the large radius limit ($R \omega_0 \to \infty$), 
Cohen \textit{et al.} derived the $Q$ ball decay rate for the Yukawa interaction~\cite{evap}. 
The scalar field configuration was taken to be the step function as 
\begin{equation}
\phi(r )=\phi_0 \theta (R -r) 
\equiv 
\begin{cases}
\phi_0, &\qquad 0 < r \leq R, \\
0, &\qquad R < r.
\end{cases}
\end{equation}
In the limit of $g \phi_0 /\omega_0 \gg 1$, 
the production rates are saturated by the Pauli exclusion principle and written as 
\begin{equation}
\frac{d}{dt}N_i = 
\left( \frac{dN}{dt} \right)_{\mathrm{sat}} 
\equiv \frac{\omega_0^3 R^2}{24 \pi}, 
\qquad \mathrm{for}\  \frac{g \phi_0}{\omega_0} \gg 1, 
\label{Cohen saturate}
\end{equation}
where $i=\chi$, $\eta$. 
On the other hand, in the limit of $g \phi_0 /\omega_0 \ll 1$, they showed that the production rates are 
\begin{equation}
\frac{d}{dt}N_i \simeq 3 \pi \left( \frac{g \phi_0}{\omega_0} \right) \times \left( \frac{dN}{dt} \right) _{\mathrm{sat}}, 
\qquad \mathrm{for}\ \frac{g \phi_0}{\omega_0} \ll 1,\  R\omega_0 \to \infty. 
\label{Cohen un sat}
\end{equation}
We can see the physical meaning of this behavior in the following way. 
The penetration length of incoming waves inside the $Q$ ball is $\sim 1/ (g\phi_0)$~\cite{evap}, 
so the effective volume of the interaction near the surface of the $Q$ ball is 
$V_{\mathrm{eff}} \sim 4 \pi R^2/ (g \phi_0)$. 
Then, the $Q$ ball decay rate, 
which is roughly the decay rate of the scalar field ($\sim g^2 \omega_0$) 
times the charge density times the effective volume, 
is $\sim (g^2 \omega_0) \times (\omega_0 \phi_0^2 )\times V_{\mathrm{eff}}$. 

\subsection{\label{sec3-2}Q ball decay rate for $R\omega_0 \sim 1$ in the Yukawa theory}
\begin{figure}[htbp]
\begin{tabular}{c}
 \includegraphics[width=100mm]{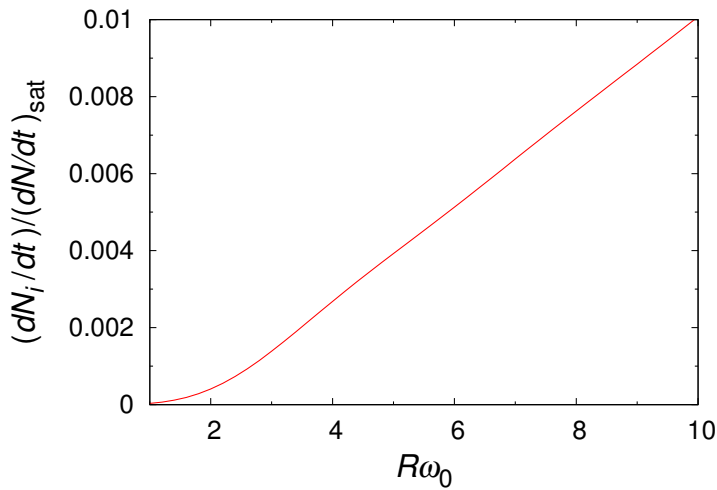}
\end{tabular}
\vspace{-0.3cm}
\caption{Production rate from the step-function type of $Q$ ball as a function of $R\omega_0$ 
with $g\phi_0 /\omega_0 = 0.01$ in the Yukawa theory. 
The vertical axis is normalized by the saturated rate of Eq.~(\ref{Cohen saturate}). 
This result can be fitted as 
$(dN_i/dt) \simeq 1.2\times 10^{-3}(R\omega_0-1.9) \times (dN/dt)_{\mathrm{sat}}$ 
for $R\omega_0 \gtrsim 2$. 
The production rates of $i=\chi$ and $\eta$ are the same. 
}
\label{Yukawa r dep}
\end{figure}

\begin{figure}[htbp]
\begin{tabular}{c}
 \includegraphics[width=100mm]{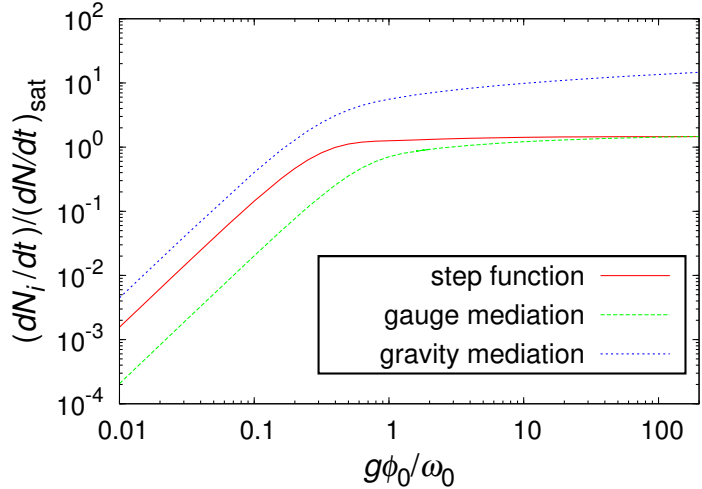}
\end{tabular}
\vspace{-0.3cm}
\caption{Production rates from $Q$ ball as a function of $g\phi_0 /\omega_0$ 
with $R\omega_0=\pi$ in the Yukawa theory. 
The types of scalar field configurations are taken to be the step-function type (red line), 
the gauge-mediation type (green dashed line) and the gravity-mediation type (blue dotted line). 
The vertical axis is normalized by the saturated rate of Eq.~(\ref{Cohen saturate}) with $R\omega_0=\pi$. 
The ratio of the step-function type of $Q$ ball to the gauge-mediation type of one to the gravity-mediation type of one 
on the production rate is 1 to 1/7 to 3 for $g\phi_0/\omega_0 \ll 1$. 
The production rate for the gravity-mediation type of $Q$ ball can be approximated as 
$(dN_i/dt) \simeq (5.9+1.75\log(g\phi_0/\omega_0)-0.02(\log(g\phi_0/\omega_0))^2 ) \times (dN/dt)_{\mathrm{sat}}$ 
for $g\phi_0/\omega_0 \gg 1$. 
The production rates of $i=\chi$ and $\eta$ are the same. 
}
\label{result Yukawa}
\end{figure}

In the realistic models of the previous section, 
the radius of the $Q$ ball $R$ is $\sim 1/ \omega_0$, 
and the effective volume of interaction become the whole region of the $Q$ ball, 
$V_{\mathrm{eff}} \sim 4 \pi R^3 / 3$~\cite{HNO}. 
The parameter dependences of the $Q$ ball decay rate are, therefore, quite different from the previous subsection. 
We derive a fitting formula for the decay rate in the limit of $g \phi_0 /\omega_0 \ll 1$ and $R \omega_0 \sim 1$. 
The decay rate are shown in Figs.~\ref{Yukawa r dep}~and~\ref{result Yukawa} 
as a function of $R$ and $g\phi_0/\omega_0$, respectively, and can be fitted as 
\begin{equation}
\left( \frac{dN_i}{dt} \right)_{\mathrm{step\ function}} \simeq 12 
\left( \frac{g \phi_0}{\omega_0} \right)^2 \left( R\omega_0 -1.9 \right) \times 
\left( \frac{dN}{dt} \right)_{\mathrm{sat}}, 
\qquad \mathrm{for}\  \frac{g\phi_0}{\omega_0} \ll 1,\  
R\omega_0 \gtrsim 2, 
\label{rate small phi}
\end{equation}
for the step-function type of scalar field configuration. 

We numerically calculate the fermion production rates 
for not only the step-function type of $Q$ ball configuration but also 
the two types of $Q$ ball configuration introduced in Sec.~\ref{sec2}; 
the numerical configuration shown in Fig.~\ref{GMSBphi} for the GMSB model and 
$\phi( r)=\phi_0 e^{-r^2/2R^2}$ for the gravity mediated SUSY breaking model. 
The results are shown in Fig.~\ref{result Yukawa} as a function of $g\phi_0/\omega_0$ with $R\omega_0=\pi$. 
We can explain the differences among the production rates for the each type of scalar field configuration. 
In the limit of $g\phi_0/\omega_0 \ll 1$, $Q$ ball charges should be taken into consideration 
because $Q$ ball decay can be regarded as a collection of decay of the scalar field. 
Since the charge density is given by $2\omega_0 \phi_0^2$ at the center of the $Q$ ball, 
the $Q$ ball charge is roughly estimated as $(8 \pi/3) R^3 \omega_0 \phi_0^2$. 
However, for the realistic configuration of the $Q$ ball, this estimation is not a good approximation. 
The ratio of the actual total charge $Q$ to the rough estimation is 
\begin{equation}
\frac{3Q}{8 \pi \omega_0 R^3 \phi_0^2} 
\simeq
\begin{cases}
1, &\qquad \mathrm{for \ step \ function}, \\[3mm]
\displaystyle{ 
\frac{3}{2 \pi^2}}, &\qquad \mathrm{for\ gauge\ mediation}, \\[3mm]
\displaystyle{ 
\frac{3 \sqrt{\pi}}{4}}, &\qquad \mathrm{for\ gravity\ mediation}, 
\end{cases}
\end{equation}
for $R\omega_0=\pi$. 
Thus, we can approximate the production rates for each type of configuration as 
\begin{equation}
\frac{d}{dt}N_i
\simeq
\begin{cases}
\displaystyle{ 
\frac{3}{2 \pi^2}}\times \left( \frac{dN_i}{dt} \right)_\mathrm{step\ function}, &\qquad \mathrm{for\ gauge\ mediation}, \\[4mm]
\displaystyle{ 
\frac{3 \sqrt{\pi}}{4}}\times \left( \frac{dN_i}{dt} \right)_\mathrm{step\ function}, &\qquad \mathrm{for\ gravity\ mediation}, 
\end{cases}
\end{equation}
for $g\phi_0/\omega_0 \ll 1$ and $R\omega_0=\pi$. 
On the other hand, for $g\phi_0/\omega_0 \gg 1$, 
the $Q$ ball decay rates are saturated by the Pauli exclusion principle. 
Since the penetration length of incoming waves inside the $Q$ ball is $\sim 1/ (g\phi)$, 
the $Q$ ball decay rates depend on $R'$, where $R'$ is determined by $g\phi(R' ) /\omega_0 \sim 1$. 
Thus, from Fig.~\ref{GMSBphi}, the decay rate for the gauge-mediation type of $Q$ ball becomes 
the same as that for the step-function type of $Q$ ball in the limit of $g\phi_0/\omega_0 \gg 1$. 
Since the gravity-mediation type of configuration is $\phi( r)=\phi_0 e^{-r^2/2R^2}$, 
the effective radius logarithmically increases as $R' \simeq R(2\log(g\phi_0/\omega_0))^{1/2}$ 
and the decay rate for that of $Q$ ball also increases as 
$(dN/dt)_\mathrm{sat}|_{R \to R'}  \simeq 2 \log(g\phi_0/\omega_0) \times (dN/dt)_\mathrm{sat}$ 
in the limit of $g\phi_0/\omega_0 \gg 1$. 
These considerations explain the decay rate of Fig.~\ref{result Yukawa}. 
The agreement on the decay rate with the step-function type of $Q$ ball is 
very good for the gauge-mediation type of one. 
On the other hand, the agreement is not so good for the gravity-mediation type of $Q$ ball. 
This disagreement comes from the fact that the step-function type and gauge-mediation type of $Q$ ball 
are thin wall configurations, but the gravity-mediation type of one is a thick wall configuration. 
We conclude that the decay rate for the gauge-mediation type of $Q$ ball can be approximated as 
\begin{equation}
\left( \frac{dN_i}{dt}\right)_{\mathrm{gauge\ mediation}} 
\simeq
\begin{cases}
\displaystyle{ 
\frac{3}{2 \pi^2}\times \left( \frac{dN_i}{dt} \right)_\mathrm{step\ function}}, 
&\qquad \displaystyle{ \mathrm{for\ }\frac{g\phi_0}{\omega_0} \ll 1 }, \\[4mm]
\displaystyle{ 
 \left( \frac{dN}{dt} \right)_\mathrm{sat}}, 
&\qquad \displaystyle{ \mathrm{for\ }\frac{g\phi_0}{\omega_0} \gg 1}, 
\end{cases}
\label{GMSB rate}
\end{equation}
where $(dN/dt)_\mathrm{sat}$ is given by Eq.~(\ref{Cohen saturate}), 
and $(dN_i/dt)_\mathrm{step\ function}$ can be fitted as Eq.~(\ref{rate small phi}). 
The decay rate for the gravity-mediation type of $Q$ ball can be fitted as 
\begin{equation}
\left( \frac{dN_i}{dt}\right)_{\mathrm{gravity\ mediation}} 
\simeq
\begin{cases}
\displaystyle{ 
3 \times \left( \frac{dN_i}{dt} \right)_\mathrm{step\ function}}, 
&\qquad \displaystyle{ \mathrm{for\ }\frac{g\phi_0}{\omega_0} \ll 1} , \\[4mm]
\displaystyle{ 
\left( 5.9+1.75\log \left( \frac{g\phi_0}{\omega_0} \right)- 0.02\left[ \log \left( \frac{g\phi_0}{\omega_0} \right) \right]^2
 \right) \times \left( \frac{dN}{dt}\right)_{\mathrm{sat}}}, 
&\qquad \displaystyle{ \mathrm{for\ }\frac{g\phi_0}{\omega_0} \gg 1}, 
\end{cases}
\label{grav rate}
\end{equation}
for $R\omega_0=\pi$ from our numerical calculation. 

\section{\label{sec4}Q ball decay rate into goldstinos}
We apply the method reviewed in the previous section to derive the goldstino production rate. 
When global SUSY is spontaneously broken, there is a goldstino 
and its interaction with a chiral multiplet at low energy is given by 
\begin{equation}
\mathcal{L}_{\mathrm{int}} = -\frac{1}{\langle F \rangle} \left( \eta 
\partial_\mu \left( 
\left( \sigma^\nu \Bar{\sigma}^\mu \chi \right)  \partial_\nu \phi^* \right)   
+ \partial_\mu \left( \partial_\nu \phi \chi^\dagger \Bar{\sigma}^\mu \sigma^\nu \right)  
\eta^\dagger \right),
\label{goldstino}
\end{equation}
where $\langle F \rangle$ is the SUSY breaking F term, and $\sigma^\mu=( \bf{1}, \sigma^i)$. 
Here, $\eta$, $\chi$ and $\phi$ are the goldstino, the chiral fermion and its superpartner, respectively. 
We assign the global $U(1)$ charge for $\phi, \chi$ and $\eta$ such as 1, 1 and 0, respectively, 
and treat the scalar field as the background field $\phi = \phi(r ) e^{-i\omega_0 t}$. 

The analysis is the same with the previous section 
once we replace the Heisenberg equations of motion of Eq.~(\ref{EOM Yukawa}) with 
\begin{equation}
\begin{cases}
\displaystyle{
i \Bar{\sigma}^{\mu}\partial_{\mu} \chi - \frac{2}{\langle F \rangle} \partial_\mu \phi \partial^\mu \eta^\dagger 
- \frac{1}{\langle F \rangle} \partial_\nu \phi \Bar{\sigma}^\nu \sigma^\mu \partial_\mu \eta^\dagger =0}, \\[4mm]
\displaystyle{
i {\sigma}^\mu \partial_\mu \eta^\dagger +\frac{1}{\langle F \rangle} \left( \partial^2 \phi^* \right)  \chi 
- \frac{1}{\langle F \rangle} \left(  \sigma^\nu \Bar{\sigma}^\mu \partial_\mu \chi \right) \partial_\nu \phi^* =0}. 
\end{cases}
\label{EOMdiff}
\end{equation}
With use of Eq.~(\ref{mode dif}), this equations are rewritten as 
\begin{equation}
\begin{cases}
\displaystyle{
\left( k_+ -i {\bf \sigma \cdot \nabla} \right)  \chi }  \\ 
\displaystyle{
\qquad+ \frac{1}{\langle F \rangle}
\left( \omega_0 \phi k_- -2\phi' \partial_r + \omega_0 \phi \left( i {\bf \sigma \cdot \nabla} \right)  
+\phi' k_- \left( i \Hat{\sigma}_r \right) - \phi' \left( i \Hat{\sigma}_r \right)  
\left( i {\bf \sigma \cdot \nabla} \right) \right) \eta^\dagger =0}, \\[4mm]
\displaystyle{
\left( -k_- +i {\bf \sigma \cdot \nabla} \right) \eta^\dagger} \\ 
\displaystyle{
\qquad+ \frac{1}{\langle F \rangle} \left( \omega_0^2 \phi +
\frac {1}{r^2} \frac{\partial }{\partial r} \left( r^2 \phi' \right) -\omega_0\phi k_+ + \omega_0  
\phi \left( i {\bf \sigma \cdot \nabla} \right) +k_+ \phi' \left( i \Hat{\sigma}_r \right) 
-\phi' \left( i \Hat{\sigma}_r \right) \left( i {\bf \sigma \cdot \nabla} \right) \right)  \chi =0}, 
\end{cases}
\label{EOMdif2}
\end{equation}
where $\Hat{\sigma}_r \equiv {\bf \sigma \cdot r} /r$ and $\phi' \equiv \partial \phi(r ) / \partial r$. 
These equations correspond to Eq.~(\ref{EOM Yukawa 2}) in the previous section. 
We need the following relation, in addition to Eq.~(\ref{formula1}), to simplify the equations: 
\begin{equation}
\Hat{\sigma}_r \Phi \left( j,m,j \pm 1/2 \right) = \Phi \left( j,m,j \mp 1/2 \right). 
\end{equation}

In this section, we use the two types of the $Q$ ball configuration $\phi(r )$ introduced in Sec.~\ref{sec2}; 
the numerical configuration shown in Fig.~\ref{GMSBphi} for the GMSB model and 
$\phi( r)=\phi_0 e^{-r^2/2R^2}$ for the gravity mediated SUSY breaking model. 

\subsection{\label{sec4-1}Goldstino production rate from the $Q$ ball 
in gauge mediation}
\begin{figure}[htbp]
\begin{tabular}{c}
 \includegraphics[width=100mm]{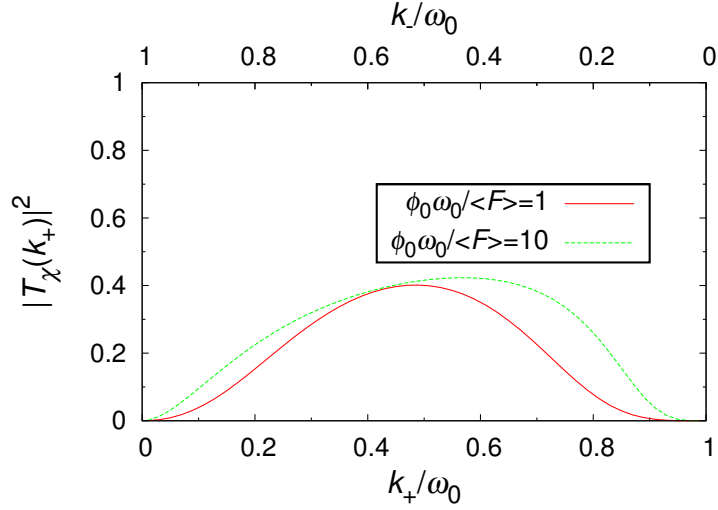}
\end{tabular}
\vspace{-0.3cm}
\caption{Energy spectrums of the outgoing $\chi$ waves $ \vert T_\chi (k_+,j) \vert^2$ 
for $\phi \omega_0 / \langle F \rangle =1$ (red line) and $\phi \omega_0  / \langle F \rangle=10$ (green dashed line) 
with $j=1/2$ in the GMSB. 
}
\label{exGMSB}
\end{figure}

In the GMSB model, 
the scalar field configuration is taken as the numerical configuration shown in Fig.~\ref{GMSBphi}. 
We show an example of the energy spectrum of outgoing $\chi$ waves in Fig.~\ref{exGMSB}. 
Using $\vert T_\chi(k_-,j) \vert^2= \vert T_\eta(k_+,j) \vert^2$ of Eq.~(\ref{dif unitarity}), along with 
$k_- \equiv \omega_0 - k_+$, we can also see the $\eta$ production rate from this figure. 
The energy dependence of the production rate is slightly asymmetric by the replacement of $k_+ \to \omega_0 -k_+$, 
so either $\chi$ or $\eta$ gets more energy from $Q$ ball than the other. 

\begin{figure}[htbp]
\begin{tabular}{c}
 \includegraphics[width=100mm]{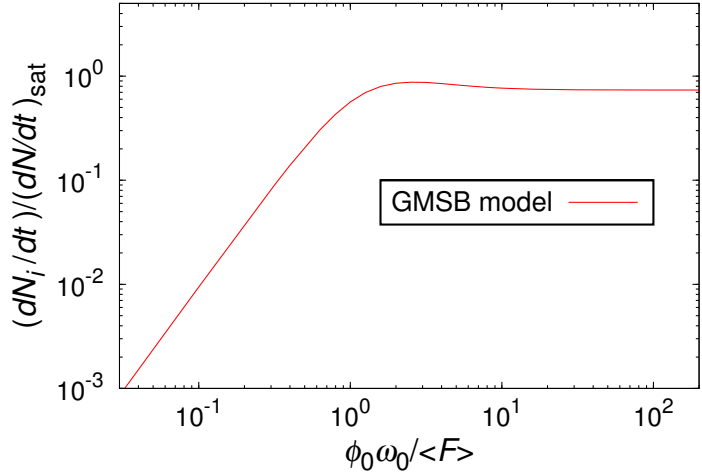}
\end{tabular}
\vspace{-0.3cm}
\caption{
Production rate from the gauge-mediation type of $Q$ ball interacting with the goldstino 
as a function of $\phi_0 \omega_0 / \langle F \rangle$. 
The vertical axis is normalized by the saturated rate of Eq.~(\ref{Cohen saturate}) with $R\omega_0=\pi$. 
The production rates of $i=\chi$ and $\eta$ are the same. 
}
\label{resultGMSB}
\end{figure}

The production rate is shown in Fig.~\ref{resultGMSB} as a function of $\phi_0 \omega_0 / \langle F \rangle$ 
and it can be written as 
\begin{equation}
\frac{d}{dt}N_i 
\simeq 
\begin{cases}
\displaystyle{
0.9 \left( \frac{ \phi_0 \omega_0 }{\langle F \rangle } \right)^2 \times 
\left( \frac{dN}{dt} \right)_{\mathrm{sat}}}, 
\qquad &\mathrm{for}\  \displaystyle{\frac{\phi_0\omega_0}{ \langle F \rangle} \ll 1}, \\[4mm]
\displaystyle{
0.74 \times \left( \frac{dN}{dt} \right)_{\mathrm{sat}}}, 
\qquad &\mathrm{for}\  \displaystyle{\frac{\phi_0\omega_0}{ \langle F \rangle} \gg 1}, 
\end{cases} 
\qquad (i=\chi, \eta), 
\label{result GMSB diff}
\end{equation}
where $(dN/dt)_{\mathrm{sat}}$ is given by Eq.~(\ref{Cohen saturate}). 
We can also get this behavior using Eqs.~(\ref{rate small phi})~and~(\ref{GMSB rate}) 
if we naively estimate derivatives in the interaction term as 
$\partial / \partial r \sim 1/R$ and $\partial / \partial t \sim \omega_0$ and use $R \omega_0 \simeq \pi$. 

\subsection{\label{sec4-2}Goldstino production rate from the $Q$ ball 
in gravity mediation}
\begin{figure}[htbp]
\begin{tabular}{c}
 \includegraphics[width=100mm]{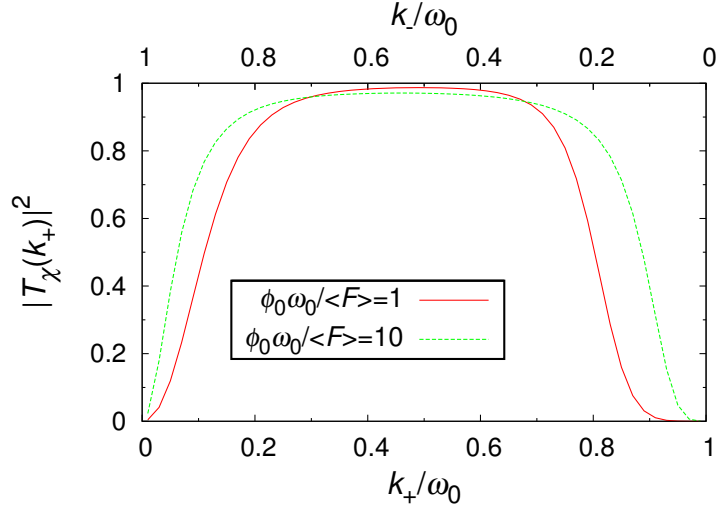}
\end{tabular}
\vspace{-0.3cm}
\caption{
Energy spectrums of the outgoing $\chi$ waves $ \vert T_\chi (k_+,j) \vert^2$ 
for $\phi \omega_0 / \langle F \rangle =1$ (red line) and $\phi \omega_0 / \langle F \rangle=10$ (green dashed line) 
with $j=1/2$ and $R\omega_0=1/ \sqrt{0.1}$ in the gravity-mediated SUSY breaking. 
}
\label{exgrav}
\end{figure}

In the gravity-mediated SUSY breaking model, the scalar field configuration is taken as 
$\phi( r)=\phi_0 \exp \left( - r^2/ 2R^2 \right) $, where $R = \vert K \vert^{-1/2} \omega_0^{-1}$. 
We show an example of the energy spectrum of outgoing $\chi$ waves in Fig.~\ref{exgrav}. 
The production rate is saturated for $k_+ \sim \omega_0/2$ by the Pauli exclusion principle. 
Using $\vert T_\chi(k_-,j) \vert^2= \vert T_\eta(k_+,j) \vert^2$ of Eq.~(\ref{dif unitarity}), along with 
$k_- \equiv \omega_0 - k_+$, 
we can also see the $\eta$ production rate from this figure. 

\begin{figure}[htbp]
\begin{tabular}{c}
 \includegraphics[width=100mm]{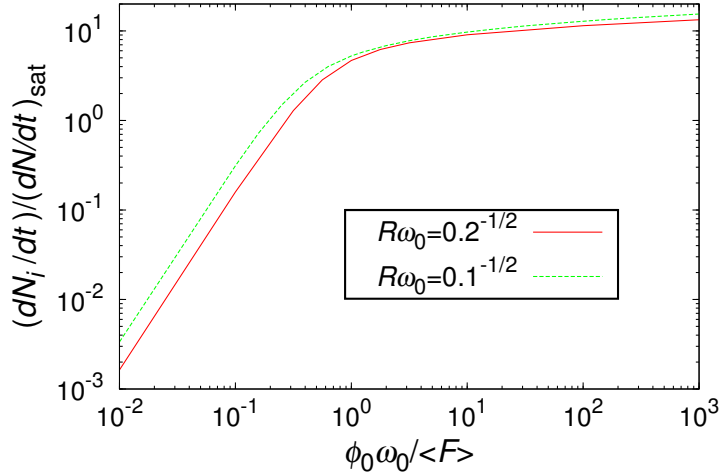}
\end{tabular}
\vspace{-0.3cm}
\caption{
Production rate from the gravity-mediation type of $Q$ ball interacting with the goldstino 
as a function of $\phi_0 \omega_0 / \langle F \rangle$ 
for $R\omega_0=1/ \sqrt{0.2}$ (red line) and $R\omega_0=1/ \sqrt{0.1}$ (green dashed line). 
The vertical axis is normalized by the saturated rate of Eq.~(\ref{Cohen saturate}). 
This result can be fitted as 
$(dN_i/dt) \simeq (6.3 +1.6 \log(\phi_0 \omega_0 / \langle F \rangle) 
-0.04(\log(\phi_0 \omega_0 / \langle F \rangle))^2) \times (dN/dt)_{\mathrm{sat}}$ 
for $R\omega_0 =1/\sqrt{0.1}$. 
The production rates of $i=\chi$ and $\eta$ are the same. 
}
\label{resultgrav}
\end{figure}

\begin{figure}[htbp]
\begin{tabular}{c}
 \includegraphics[width=100mm]{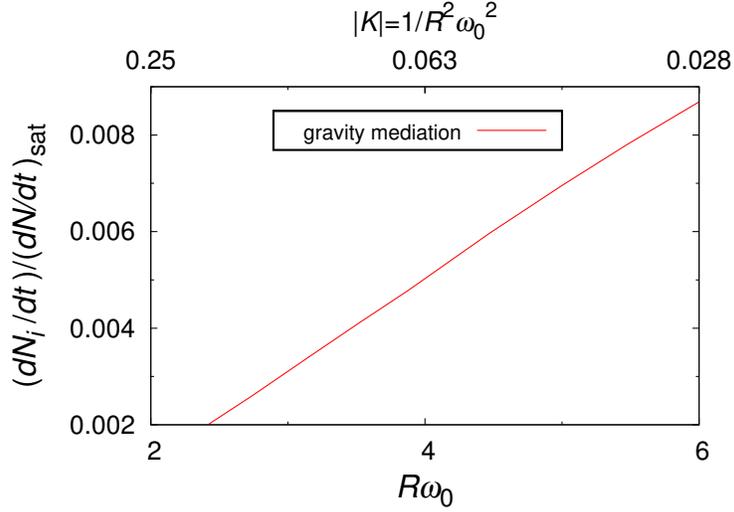}
\end{tabular}
\vspace{-0.3cm}
\caption{
Production rate from the gravity-mediation type of $Q$ ball interacting with the goldstino 
as a function of $R\omega_0$ with $\phi_0 \omega_0/\langle F \rangle = 0.01$. 
The vertical axis is normalized by the saturated rate of Eq.~(\ref{Cohen saturate}). 
This result can be fitted as 
$(dN_i/dt) \simeq 1.9\times 10^{-3} (R\omega_0-1.3) \times (dN/dt)_{\mathrm{sat}}$. 
The production rates of $i=\chi$ and $\eta$ are the same. 
}
\label{diff r dep}
\end{figure}

The production rates are shown in Figs.~\ref{resultgrav}~and~\ref{diff r dep} 
as a function of $\phi_0 \omega_0 / \langle F \rangle$ and $R\omega_0$, respectively, 
and they can be written as 
\begin{equation}
\frac{d}{dt}N_i 
\simeq 
\begin{cases}
\displaystyle{ 
19 (R\omega_0 -1.3)
\left( \frac{\phi_0 \omega_0}{ \langle F \rangle} \right)^2 \times 
\left( \frac{dN}{dt} \right)_{\mathrm{sat}}}, 
\qquad &\mathrm{for}\ \displaystyle{ \frac{\phi_0\omega_0}{ \langle F \rangle} \ll 1}, \\[4mm]
\displaystyle{ 
\left( 6.3+1.6\log \left( \frac{g\phi_0}{\omega_0} \right)-0.04\left[ \log \left( \frac{g\phi_0}{\omega_0} \right) \right]^2 \right) 
\times  \left( \frac{dN}{dt} \right)_{\mathrm{sat}}}, 
\qquad &\mathrm{for}\ \displaystyle{ \frac{\phi_0\omega_0}{ \langle F \rangle} \gg 1, \ R\omega_0=\frac{1}{\sqrt{10}}}, 
\end{cases}.
\label{result grav diff}
\end{equation}
We can also get this behavior using Eqs.~(\ref{rate small phi})~and~(\ref{grav rate}) 
if we naively estimate derivatives in the interaction term as 
$\partial / \partial r \sim 1/R$ and $\partial / \partial t \sim \omega_0$ and use $\omega_0 \gtrsim 1/R$. 

\section{\label{sec5}Q ball decay with a massive fermion}
In this section, we consider the Yukawa interaction with a massive fermion. 
We are interested in this case because squarks 
interact with quarks and gluinos which have large masses we can not ignore~\cite{KLS1}. 
The Lagrangian density is written as 
\begin{equation}
\mathcal{L} = 
\chi^\dagger i \Bar{ \sigma }^\mu \partial_\mu \chi 
+\eta^\dagger i \Bar{ \sigma }^\mu \partial_\mu \eta
- g \phi^*  \chi \eta -g \phi \chi^\dagger \eta^\dagger 
-\frac{1}{2} M \eta \eta -\frac{1}{2} M \eta^\dagger \eta^\dagger.
\end{equation}
We assign the global $U(1)$ charge for $\phi, \chi$ and $\eta$ such as 1, 1 and 0, respectively.
The background field is $\phi = \phi(r ) e^{-i\omega_0 t}$. 
In this section, we take the following background field configuration for simplicity: 
\begin{equation}
\phi(r ) = \phi_0 \theta (R -r) 
\equiv 
\begin{cases}
\phi_0, &\qquad 0 < r \leq R, \\
0, &\qquad R < r.
\end{cases}
\end{equation}

Inside the $Q$ ball, the Heisenberg equations of motion are 
\begin{equation}
\begin{cases}
i \Bar{\sigma}^{\mu}\partial_{\mu} \chi - g \phi \eta^\dagger =0, \\
i {\sigma}^\mu \partial_\mu \eta^\dagger - g \phi^* \chi - M \eta =0, \\
i {\sigma}^{\mu}\partial_{\mu} \chi^\dagger - g \phi^* \eta =0, \\
i \Bar{\sigma}^\mu \partial_\mu \eta - g \phi \chi^\dagger - M \eta^\dagger =0. 
\end{cases}
\label{EOMmass}
\end{equation}
We want to obtain solutions of these equations by the mode and the angular momentum expansion. 
The symmetry of the simultaneous time translations and $U(1)$ rotations allows us to expand the solution as 
\begin{equation}
\begin{cases}
\chi \propto e^{-i\omega  t} \equiv e^{-ik_+ t }, \\
\chi^\dagger \propto e^{-i \left( \omega-2\omega_0\right) t} \equiv e^{ik_- t }, \\
\eta \propto e^{-i(\omega -\omega_0) t } \equiv e^{-ik_\eta t }, \\
\eta^\dagger \propto e^{-i(\omega -\omega_0) t } \equiv e^{-ik_\eta t }, 
\label{massive expansion}
\end{cases}
\end{equation}
where we include $\chi^\dagger$ and $\eta$ as well as $\chi$ and $\eta^\dagger$ 
because they mix with each other through the mass term (see Eq.~(\ref{EOMmass})). 
We derive the fermion production rate through the Bogoliubov transformation 
between creation and annihilation operators at $t \to \pm \infty$. 
Thus, we restrict our attention to $0 < \omega < 2\omega_0$. 

Now, the solutions inside the $Q$ ball can be written by the spinor solutions in the following form: 
\begin{equation}
\begin{cases}
\chi = A e^{-ik_+ t} u(k,j,m; {\bf r}), \\
\chi^\dagger =B e^{ik_- t} u(k,j,m; {\bf r}), \\
\eta =C e^{-ik_\eta t} u(k,j,m; {\bf r}), \\
\eta^\dagger =D e^{-ik_\eta t} u(k,j,m; {\bf r}), 
\end{cases}
\end{equation}
where $A, B, C, D$ and $k$ are some constants fixed by solving Eq.~(\ref{EOMmass}) up to an overall normalization. 
When we substitute these into Eq.~(\ref{EOMmass}), we find 
\begin{equation}
\begin{cases}
\left( k_+ + k \right) A - g\phi D =0, \\
\left( k_\eta - k \right) D - g\phi A - M C =0, \\
\left( -k_- - k \right) B - g\phi C =0, \\
\left( k_\eta + k \right) C - g\phi B -M D =0. 
\end{cases}
\end{equation}
When we eliminate $A, B, C$ and $D$ from these equations, we find that $k$ obeys 
\begin{equation}
((k+k_+)(k-k_\eta)+g^2\phi_0^2) ((k+k_-)(k+k_\eta)
+g^2\phi_0^2) +M^2(k+k_+)(k+k_-)=0. 
\end{equation}
Thus, recalling that $k_+, k_-$ and $k_\eta$ are fixed in Eq.~(\ref{massive expansion}), 
we obtain four independent solutions inside the $Q$ ball. 
After matching the interior and exterior solutions at $r=R$, 
we can write annihilation and creation operators of outgoing waves 
in terms of annihilation and creation operators of incoming waves 
in the same way as the previous section. 

As mentioned above, we are interested in the case that the scalar field interacts with heavy gluinos. 
Typically, $\omega_0$ is GeV range in the GMSB, and the mass of gluino is TeV range, so $M \gg \omega_0$. 
Fortunately, in the case of $M > \omega_0$, we can calculate the fermion production rate very easily. 
However, we also present the calculation in the case of $M < \omega_0$. 

\subsection{\label{sec5-1}Case of $M > \omega_0$}
In the case of $M> \omega_0$, 
coefficients $T$ and $R$ of $\eta$ are irrelevant since $\eta$ has no degree of freedom outside the $Q$ ball, 
and the boundary condition is $\eta \to 0$ as $r \to \infty$. 
So, we need to get the coefficients of only $\chi$ and $\chi^\dagger$, 
and the analysis is almost equivalent to the previous section. 
In this case, however, $\chi$ and $\chi^\dagger$ are related to each other by hermitian conjugation. 
Thus, the Bogoliubov transformation can be written as 
\begin{equation}
\begin{pmatrix}
a_\chi (k_+,j,m) \\[2mm]
(-1)^{m_-} a^\dagger_\chi (k_-,j,-m) \\
\end{pmatrix} _{\mathrm{out}} 
= 
\begin{pmatrix}
R^0_\chi(k_+,j)	&T_\chi^0(k_+,j) \\[2mm]
-T^{0*}_\chi(k_-,j) &R_\chi^{0*} (k_-,j) 
\end{pmatrix}
\begin{pmatrix}
a_\chi(k_+,j,m) \\[2mm]
(-1)^{m_-} a^\dagger_\chi(k_-,j,-m) \\
\end{pmatrix}_{\mathrm{in}}, 
\label{bogo large m}
\end{equation}
where $R^0_\chi$ and $T_\chi^0$ are coefficients 
fixed by matching the interior and exterior solutions with the boundary condition $\eta \to 0$ as $r \to \infty$, 
and we write the coefficients in order to maintain consistency with hermitian conjugation. 

\begin{figure}[htbp]
\begin{tabular}{c}
 \includegraphics[width=100mm]{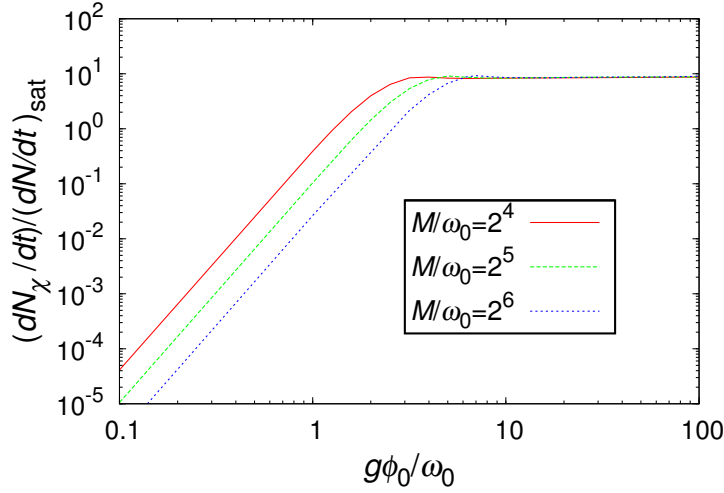}
\end{tabular}
\vspace{-0.3cm}
\caption{
Production rates of $\chi$ from $Q$ balls as a function of $g \phi_0/\omega_0$ 
for $M/\omega_0=2^4, 2^5$ and $2^6$ with $R\omega_0=\pi$ 
in the Yukawa theory with a massive fermion. 
The vertical axis is normalized by the saturated rate of Eq.~(\ref{Cohen saturate}). 
}
\label{mlarge}
\end{figure}

We can calculate the $\chi$ production rate in the same way as the previous section 
once we replace $\omega_0$ with $2\omega_0$ in Eq.~(\ref{evaporationrate}). 
The production rate is shown in Fig.~\ref{mlarge} as a function of $g \phi_0/\omega_0$, 
and it can be written as 
\begin{equation}
\frac{d}{dt} N_\chi 
\simeq 
\begin{cases}
\displaystyle{
13 \left( \frac{\omega_0}{M}\right)^2 \left( \frac{g\phi_0}{\omega_0} \right)^4 
\times \left[ \left. \left( \frac{dN}{dt} \right)_{\mathrm{sat}}\right\vert_{\omega_0 \to 2\omega_0}\right]}, 
\qquad &\mathrm{for}\ \displaystyle{ \left( \frac{\omega_0}{M} \right) 
\left( \frac{g \phi_0}{\omega_0} \right)^2 \ll 1,\ R\omega_0 =\pi}, \\[4mm]
\displaystyle{
1.1 \times 
\left[ \left. \left( \frac{dN}{dt} \right)_{\mathrm{sat}}\right\vert_{\omega_0 \to 2\omega_0}\right]}, 
\qquad &\mathrm{for}\ \displaystyle{ \left( \frac{\omega_0}{M} \right) 
\left( \frac{g \phi_0}{\omega_0} \right)^2 \gg 1,\ R\omega_0 =\pi}, 
\end{cases}
\label{resultmassive}
\end{equation}
where we replace $\omega_0$ with $2\omega_0$ in the saturated production rate of Eq.~(\ref{Cohen saturate}) 
because the energy spectrum of $\chi$ is now in the interval $(0, 2  \omega_0)$. 
We can understand the above behavior of the production rate by integrating out the heavy particle. 
Because the effective interaction after integrating out the heavy particle has the form of the 
Yukawa interaction as $\left( g^2\phi^{*2}_0 / M \right) (\chi \chi/2)$, 
the production rate is $12 (\pi-1.9) (g^2\phi_0^2/M\omega_0)^2 \times (dN/dt)_{\mathrm{sat}}$ 
from Eq.~(\ref{rate small phi}) and $R\omega_0 = \pi$. 
Thus, we conclude that 
the production rate calculated in the effective theory is consistent with our numerical result of Eq.~(\ref{resultmassive}) 
if we replace $\omega_0$ with $2\omega_0$ in the saturated production rate. 

\begin{figure}[htbp]
\begin{center}
\setlength{\tabcolsep}{3pt}
\begin{tabular}{c}
\resizebox{43mm}{!}{\includegraphics{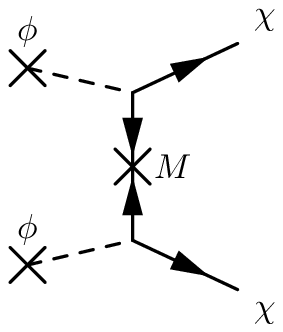}} 
\end{tabular}
\end{center}
\caption{Diagram for $\phi \phi \to \chi \chi$.}
\label{diag3}
\end{figure}

We can explain the above behavior of the production rate in another way. 
Recall Eq.~(\ref{rate small phi}) which is derived in the massless case can be interpreted as 
the decay rate $\Gamma_{\phi}$ times the charge density times the effective volume, i.e. 
$\Gamma_{\phi} \times (\omega_0 \phi_0^2) \times V_{\mathrm{eff}}$. 
On the other hand, because the reaction we consider here is two particle scattering process $\phi \phi \to q q$, 
the $Q$ ball decay rate should be estimated as 
($\mathrm{flux}) \times (\mathrm{cross\ section}) \times \omega_0 \phi_0^2 \times V_{\mathrm{eff}}$. 
The flux is the number density $\omega_0 \phi^2$ times the relative velocity 
and the cross section can be estimated as $g^4 / M^2$ from Fig.~\ref{diag3}. 
Then if we assume the relative velocity as $O$(1) 
we can get the same parameter dependences with the first line of Eq.~(\ref{resultmassive}).

\subsection{\label{sec5-2}Case of $M < \omega_0$}
In the case of $M < \omega_0$, $\eta$ can propagate outside the $Q$ ball and 
its coefficients are also important. 
Outside the $Q$ ball, the fields obey the free equations of motion. 
The free field expansion of the field $\chi$ is the same as the previous section, 
but the expansion of the field $\eta$ is complicated by the presence of the mass term. 
The modes of $\eta \propto e^{-i k_\eta t}$ and 
$\eta^\dagger \propto e^{-i k_\eta t}$ mix with each other by the equations of motion as 
\begin{equation}
\begin{cases}
-i{\bf \sigma} \cdot {\bf \nabla} \eta = - k_\eta \eta  +M \eta^\dagger, \\ 
-i{\bf \sigma} \cdot {\bf \nabla} \eta^\dagger = k_\eta \eta^\dagger - M \eta.
\end{cases}
\end{equation}
We define the following linear combinations to make $\eta_1$ and $\eta_2$ independent from each other: 
\begin{equation}
\begin{pmatrix}
\eta_1 \\[1mm]
\eta_2 
\end{pmatrix}
=
\begin{pmatrix}
-\frac{M}{A} &\frac{k_\eta  +p}{A}\\[1mm]
-\frac{k_\eta  +p}{A} &\frac{M}{A}
\end{pmatrix}
\begin{pmatrix}
\eta \\[1mm]
\eta^\dagger 
\end{pmatrix}, 
\label{eta trans}
\end{equation}
where $p \equiv \sqrt{k_\eta^2-M^2}$ and 
$A \equiv A(k_\eta) \equiv \sqrt{\left( \left( k_\eta +p \right) ^2 + M^2 \right) p / k_\eta}=\sqrt{(k_\eta + p)^2 - M^2}$. 
Then, with use of Eq.~(\ref{u hankel}), the solutions are 
\begin{equation}
\begin{cases}
\eta_1 = c_{\eta_1 \mathrm{in}} u^{(2)}(p,j,m,;{\bf r} ) e^{-ik_\eta t } 
+ c_{\eta_1 \mathrm{out}} u^{(1)}(p,j,m,;{\bf r} ) e^{-ik_\eta t }, \\
\eta_2 = c_{\eta_2, \mathrm{in}}   u^{(1)}(-p,j,m,;{\bf r} ) e^{-ik_\eta t } 
+ c_{\eta_2, \mathrm{out}} u^{(2)}(-p,j,m,;{\bf r} ) e^{-ik_\eta t }, 
\end{cases}
\end{equation}
where $c_{\eta_i}$ ($i=$1, 2) are arbitrary constants. 
When $ \vert k_\eta \vert  < M$, $p$ is pure imaginary, and 
the solutions of $\eta$ and $\eta^\dagger$ damp outside the $Q$ ball. 
This is because there is no degree of freedom for $\eta$ at an energy scale below the mass of $\eta$. 
Thus, we can write the mode expansion as 
\begin{multline}
\eta = \sum_{j,m} \int^{\infty}_M dk_\eta \left( c_{\eta_1 \mathrm{in}}(k_\eta ,j,m)e^{-ik_\eta t} 
\frac{M}{A(k_\eta)} u^{(2)}(p,j,m;{\bf r} ) \right. \\ 
\left. \left. + c_{\eta_1 \mathrm{in}}^\dagger(k_\eta ,j,-m)e^{ik_\eta t} 
(-1)^{m_+} \frac{ k_\eta +p }{A(k_\eta)} u^{(1)}(p,j,m;{\bf r}) \right) \right\vert_{p=\sqrt{k_\eta ^2-M^2}} \\
+( \mathrm{terms\ of\ } c_{\eta_2 \mathrm{in}}) 
+(\mathrm{terms\ of\ outgoing\ waves}). 
\label{eta expansion2}
\end{multline}
When we quantize the field $\eta$, 
we impose the  canonical anticommutation relations for $\eta$ and $\eta^\dagger$. 
Then, the coefficients $c_{\eta_1}$ and $c_{\eta_2}$ become operators, and 
their anticommutation relations are given by 
$\{ c_{\eta_i}^\dagger \left( k_\eta, j, m \right), c_{\eta_{i'}} \left( k_\eta', j', m' \right) \} = 
\delta \left( k_\eta -k_\eta' \right) \delta_{j j'} \delta_{m m'} \delta_{i i'}$, 
where we have used $A(k_\eta )=\sqrt{\left(  \left(  k_\eta  +p \right) ^2 + M^2 \right)  p / k_\eta }$ and 
\begin{equation}
\int_M^\infty dk_\eta \left[ \frac{k_\eta}{p}u^{(i)\dagger}(p; {\bf r} )u^{(i)}(p; {\bf r}' )+(p \to -p) \right]_{p=\sqrt{k_\eta ^2-M^2}}
= \int _{-\infty}^\infty dp\ u^{(i)\dagger}(p; {\bf r} )u^{(i)}(p; {\bf r}' ). 
\end{equation}
The operator $c_{\eta_i}(k_\eta ,j,m)$ is the annihilation operator of the energy $E=k_\eta$, 
and its normalization is the same as $a_\chi$. 

Next, we consider the whole system including the $Q$ ball background. 
We expand the solution as Eq.~(\ref{massive expansion}). 
If $k_\eta >0$ ($k_\eta <0$), 
we take $c_{\eta_1}$, $c_{\eta_2}$ ($c_{\eta_1}^\dagger$, $c_{\eta_2}^\dagger$) terms in Eq.~(\ref{eta expansion2}). 
In the case of $\vert k_\eta \vert <M$, $\eta$ has no degree of freedom outside the $Q$ ball 
and the analysis is the same as the case of the previous subsection. 
Thus, the Bogoliubov transformation can be written as Eq.~(\ref{bogo large m}). 
On the other hand, in the case of $\vert k_\eta \vert >M$, there are also incoming and outgoing $\eta$ waves, 
and so the coefficients of $\eta$ as well as $\chi$ are important. 
The terms mixing with each other are 
\begin{equation}
\begin{cases}
a_{\chi in}(k_+,j,m) e^{-ik_+ t } u^{(1)} ( -k_+, j, m ; {\bf r}) 
&+ (\mathrm{in}\to \mathrm{out,} \quad u^{(1)} \to u^{(2)}), \\ 
a_{\chi in}^\dagger (k_-,j,-m) e^{ik_- t } (-1)^{m_-} u^{(2)} ( -k_-, j, m ; {\bf r}) 
&+ (\mathrm{in}\to \mathrm{out,} \quad u^{(2)} \to u^{(1)}), \\
c_{\eta_1 in}(k_\eta,j,m) e^{-ik_\eta t }u^{(2)} ( p, j, m ; {\bf r}) 
&+ (\mathrm{in}\to \mathrm{out,} \quad u^{(2)} \to u^{(1)}), \\
c_{\eta_2 in}(k_\eta,j,m) e^{-ik_\eta t }u^{(1)} ( -p, j, m ; {\bf r}) 
&+ (\mathrm{in}\to \mathrm{out,} \quad u^{(1)} \to u^{(2)}), 
\end{cases}
\end{equation}
for $k_\eta > M$. 
After solving the Heisenberg equation of motion and matching the interior and exterior solutions, 
the outgoing waves can be written by the incoming waves. 
Thus we write the Bogoliubov transformation for the case of $k_\eta > M$ as 
\begin{equation}
\begin{pmatrix}
a_{\chi \mathrm{out}} \left( k_+,m \right)   \\[1mm]
(-1)^{m_-} a_{\chi \mathrm{out}}^\dagger \left( k_-,-m \right)   \\[1mm]
c_{\eta_1 \mathrm{out}} \left( k_\eta,m \right)   \\[1mm]
c_{\eta_2 \mathrm{out}} \left( k_\eta,m \right)   
\end{pmatrix}
=
\begin{pmatrix}
R_{\chi \chi} \left( k_+ \right)   	&T_{ \chi^\dagger \chi} \left( k_+ \right)   	
	&R_{\eta_1\chi} \left( k_+ \right)  	&R_{\eta_2 \chi} \left( k_+ \right)  	\\[1mm]
T_{\chi \chi^\dagger} \left( k_- \right)   	&R_{ \chi^\dagger  \chi^\dagger} \left( k_- \right)   
	&T_{\eta_1 \chi^\dagger} \left( k_- \right)   	&T_{\eta_2 \chi^\dagger} \left( k_- \right)   	\\[1mm]
R_{\chi \eta_1} \left( k_\eta \right)   	&T_{ \chi^\dagger \eta_1} \left( k_\eta \right)   
	&R_{\eta_1 \eta_1} \left( k_\eta \right)  	&R_{\eta_2 \eta_1} \left( k_\eta \right)  	\\[1mm]
R_{\chi \eta_2} \left( k_\eta \right)   	&T_{ \chi^\dagger \eta_2} \left( k_\eta \right)   	
	&R_{\eta_1 \eta_2} \left( k_\eta \right)  	&R_{\eta_2 \eta_2} \left( k_\eta \right)  	
\end{pmatrix} 
\begin{pmatrix}
a_{\chi\mathrm{in}} \left( k_+ ,m\right)   \\[1mm]
(-1)^{m_-} a_{\chi \mathrm{in}}^\dagger \left( k_-,-m \right)   \\[1mm]
c_{\eta_1 \mathrm{in}} \left( k_\eta ,m \right)   \\[1mm]
c_{\eta_2 \mathrm{in}} \left( k_\eta ,m \right)   
\end{pmatrix}, 
\label{matrixm}
\end{equation}
where we omitted the argument $j$. 
The coefficients $R_i$ and $T_i$ are fixed by matching the interior and exterior solutions at $r=R$ 
and do not depend on $m$ due to rotational invariance. 
The anticommutation relations imply that the matrix of the right hand side is a unitary matrix. 

We can calculate the production rates as 
\begin{equation}
\begin{split}
\frac{d}{dt} {n}_\chi \left( k_+,j,m \right)  &\equiv \frac{2\pi}{ T} \langle 0_{\mathrm{in}} \vert  
a_{\chi \mathrm{out}}^\dagger(k_+,j,m) a_{\chi \mathrm{out}}(k_+,j,m) \vert 0_{\mathrm{in}} \rangle,  \\
& = 
\begin{cases}
\vert T_{\chi^\dagger \chi} (k_+,j) \vert^2, 
& \qquad \left( M< k_\eta  \right), \\[1mm]
\vert T_\chi^0(k_+,j) \vert^2,  
& \qquad \left( \vert k_\eta  \vert < M \right), \\[1mm]
\left( \vert T_{\chi \chi^\dagger} (k_+,j) \vert^2 + \vert T_{\eta_1 \chi^\dagger}(k_+,j) \vert^2
+ \vert T_{\eta_2 \chi^\dagger} (k_+,j) \vert^2 \right),  
& \qquad \left( k_\eta < -M \right), 
\end{cases}
\end{split}
\end{equation}
for the $\chi$ waves and as 
\begin{equation}
\begin{split}
 \frac{d}{dt} {n}_{\eta_i} \left( k_\eta, j,m \right) 
&\equiv \frac{2\pi}{T} \langle 0_{\mathrm{in}} 
\vert c_{\eta_i \mathrm{out}}^\dagger(k_\eta, j,m) c_{\eta_i \mathrm{out} }(k_\eta,j,m) \vert 0_{\mathrm{in}} \rangle, \\
& = 
\begin{cases}
0, 
& \qquad \left(  k_\eta < M \right),  \\[1mm]
\vert T_{\chi^\dagger \eta_i}(k_\eta,j) \vert^2, 
& \qquad \left(  M<k_\eta < \omega_0\right), 
\end{cases}
\end{split}
\end{equation}
for the $\eta$ waves. 
Due to the unitarity of the Bogoliubov transformation, 
we can see that the production rates satisfy the Pauli exclusion principle, $n_i \leq 1$ ($i=\chi$, $\eta_1$, $\eta_2$). 
Especially, from the second column and the second row of Eq.~(\ref{matrixm}), we also have 
\begin{equation}
\vert T_{\chi \chi^\dagger}(k_-,j) \vert^2 + \vert T_{\eta_1 \chi^\dagger}(k_-,j) \vert^2 
+ \vert T_{\eta_2 \chi^\dagger} (k_-,j) \vert^2  
= \vert T_{\chi^\dagger \chi}(k_+,j) \vert^2 + \vert T_{\chi^\dagger \eta_1}(k_\eta,j) \vert^2 
+ \vert T_{\chi^\dagger \eta_2}(k_\eta,j) \vert^2, 
\end{equation}
for $M< k_\eta$. In other words, using $k_+ \equiv \omega$, $k_- \equiv 2\omega_0 -\omega$ 
and $k_\eta \equiv \omega -\omega_0$, we have 
\begin{equation}
\frac{d}{dt} {n}_\chi (E=2\omega_0 - \omega)
=
\frac{d}{dt} {n}_\chi (E=\omega) 
+ \sum_{i=1,2} \frac{d}{dt} {n}_{\eta_i} (E= \omega - \omega_0), 
\label{unitarity all}
\end{equation}
where $E$ denotes an energy of each particle and we omitted the argument $j$ and $m$. 
From this, we can understand the processes of the $Q$ ball decay as the superposition of 
\begin{equation}
\begin{cases}
\phi (E= \omega_0) + \phi (E= \omega_0) \to \chi (E= \omega)+ \chi (E=2 \omega_0-\omega),   \\ 
\phi (E= \omega_0) \to \chi (E= \omega)+ \eta_i (E= \omega_0-\omega), \qquad (i=1, 2). 
\end{cases} 
\label{elemental process}
\end{equation}
%

\begin{figure}[htbp]
\begin{tabular}{c}
 \includegraphics[width=100mm]{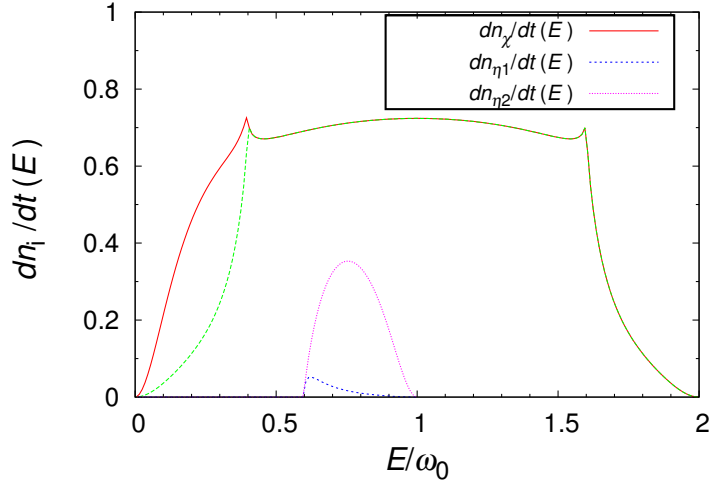}
\end{tabular}
\vspace{-0.3cm}
\caption{
Energy spectrums of the outgoing $\chi$ (red line), $\eta_1$ (blue dashed line) and $\eta_2$ (pink dotted line) waves 
with $g\phi_0/\omega_0=10$, $M/\omega_0=0.6$, $R\omega_0=\pi$ and $j=1/2$. 
The green dashed line shows $dn_\chi /dt(E)- \sum_i dn_{\eta_i} / dt (\omega_0 -E)$. 
}
\label{examplerate}
\end{figure}

\begin{figure}[htbp]
\begin{center}
\setlength{\tabcolsep}{3pt}
\begin{tabular}{l l}
\resizebox{83mm}{!}{\includegraphics{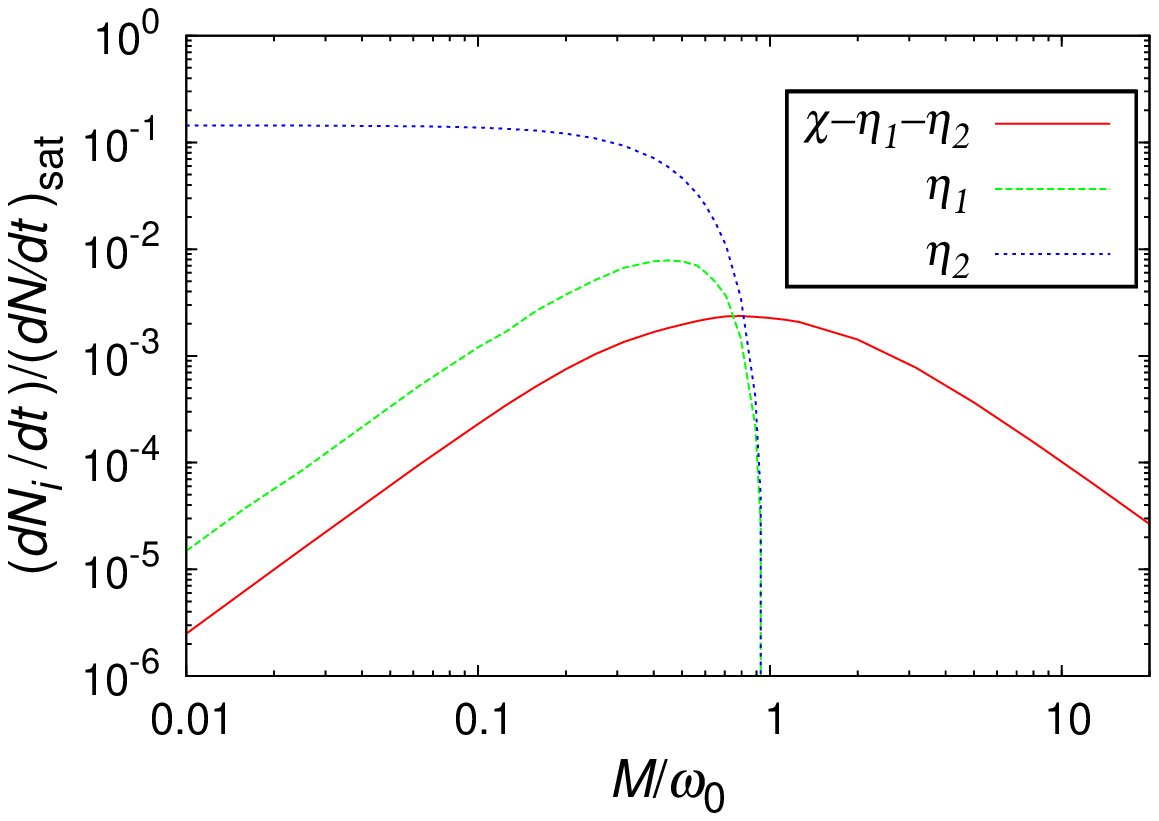}} &  
\resizebox{83mm}{!}{\includegraphics{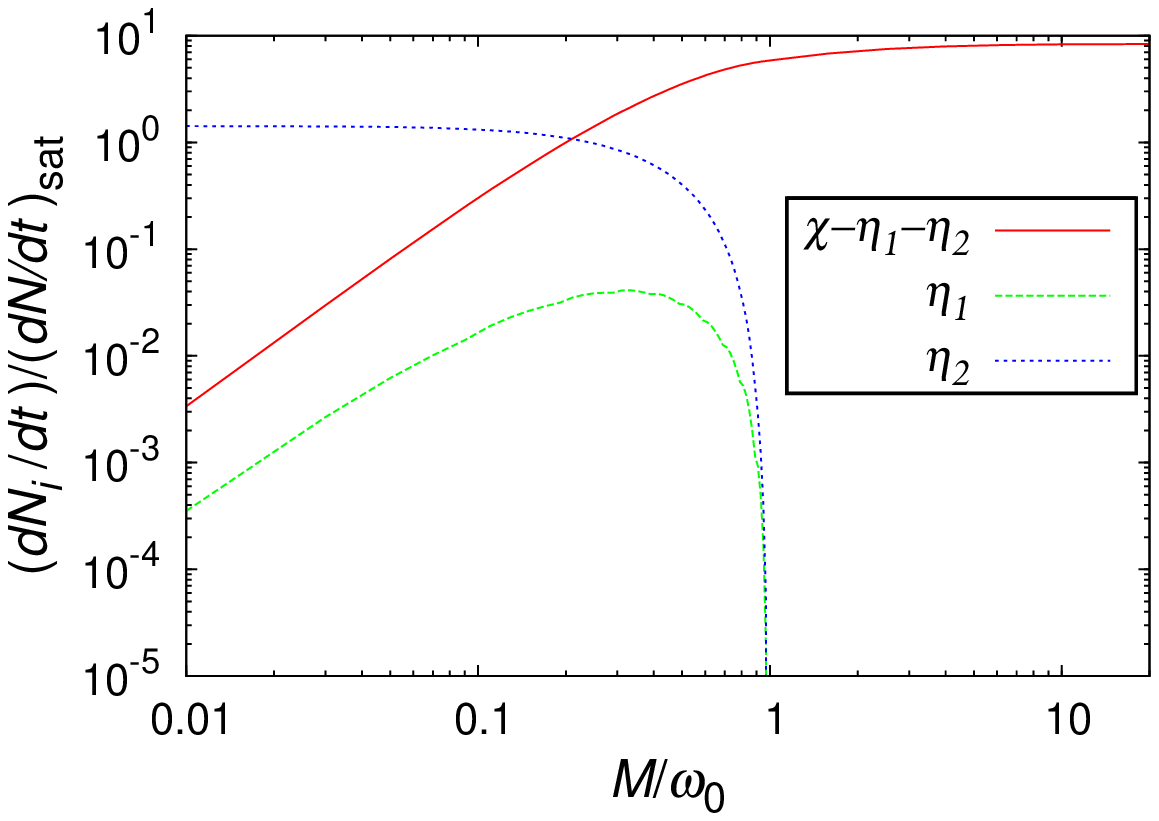}} \\
\end{tabular}
\end{center}
\caption{Production rates of $\chi$, $\eta_1$ and $\eta_2$ from $Q$ balls as a function of $M/\omega_0$ 
for $g\phi_0/\omega_0=0.1$ (left panel) and for $g\phi_0/\omega_0=10$ (right panel) 
with $R\omega_0=\pi$ in the Yukawa theory with a massive fermion. 
The vertical axis is normalized by the saturated rate of Eq.~(\ref{Cohen saturate}). 
}
\label{resultmass}
\end{figure}

\begin{figure}[htbp]
\begin{tabular}{c}
 \includegraphics[width=100mm]{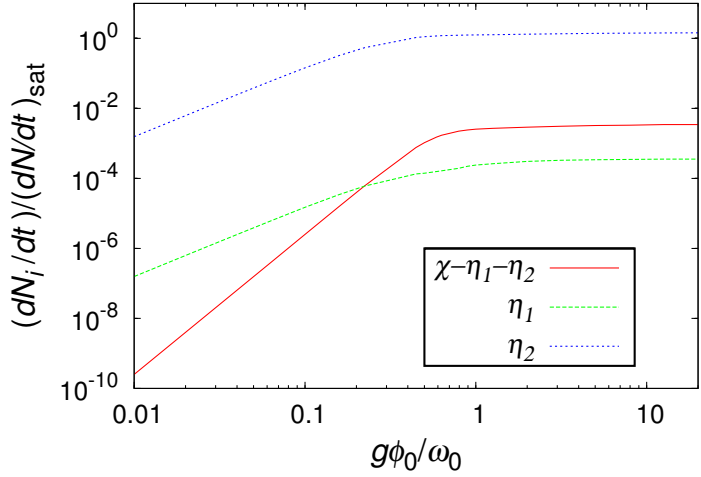}
\end{tabular}
\vspace{-0.3cm}
\caption{
Production rates of $\chi$, $\eta_1$ and $\eta_2$ from $Q$ balls as a function of $g\phi_0/\omega_0$ 
with $R\omega_0=\pi$ and $M/\omega_0=0.01$ in the Yukawa theory with a massive fermion. 
The vertical axis is normalized by the saturated rate of Eq.~(\ref{Cohen saturate}). 
}
\label{resultmass2}
\end{figure}

Fig.~\ref{examplerate} shows an example of the energy spectrum of each field. 
In this figure, $dn_\chi /dt(E)- \sum_i dn_{\eta_i} / dt (\omega_0 -E)$ is symmetrical; 
that is, the production rates satisfy Eq.~(\ref{unitarity all}). 

Figs.~\ref{resultmass}~and~\ref{resultmass2} show the production rates of each field 
as a function of $M/\omega_0$ and $g\phi_0/\omega_0$, respectively. 
We plot the $\chi$ production rate minus the $\eta$ production rate instead of the total $\chi$ production rate; 
in other words, we plot the contribution of only the first line of Eq.~(\ref{elemental process}). 
From Figs.~\ref{mlarge}~and~\ref{resultmass}, 
we can see that the production rate is proportional to $g^4\phi^4_0/M^2 \omega_0^2$ in the limit of 
$g \phi_0 /\omega_0 \ll 1$ and $M/\omega_0 \gg 1$, and it can be written as Eq.~(\ref{resultmassive}). 
From Figs.~\ref{resultmass}~and~\ref{resultmass2}, 
the production rates in the limit of $M /\omega_0 \ll 1$ and $g\phi_0/\omega_0 \gg 1$ can be read as 
\begin{equation}
\begin{cases}
\displaystyle{
\frac{d}{dt} N_\chi \simeq 4 \left( \frac{M}{ \omega_0}\right)^2 
\times \left[ \left. \left(\frac{dN}{dt}\right)_{\mathrm{sat}}\right\vert_{\omega_0 \to 2\omega_0} \right]}, \\[4mm]
\displaystyle{
\frac{d}{dt} N_{\eta_1} \simeq 3 \left(\frac{M}{\omega_0}\right)^2 \times \left(\frac{dN}{dt}\right)_{\mathrm{sat}}}, \\[4mm]
\displaystyle{
\frac{d}{dt} N_{\eta_2} \simeq 1.4 \times \left(\frac{dN}{dt}\right)_{\mathrm{sat}}}, 
\end{cases}
\qquad \mathrm{for}\ \displaystyle{ \frac{M }{\omega_0} \ll 1,\  \frac{g\phi_0}{\omega_0} \gg 1,\  R\omega_0 =\pi}. 
\label{low mass result large phi}
\end{equation}
From Figs.~\ref{resultmass}~and~\ref{resultmass2}, 
the production rates in the limit of $M/\omega_0 \ll 1$ and $g \phi_0/\omega_0 \ll 1$ can be read as 
\begin{equation}
\begin{cases}
\displaystyle{
\frac{d}{dt} N_\chi  \simeq 31 \left( \frac{M}{ \omega_0}\right)^2 \left(\frac{g \phi_0}{\omega_0}\right)^4 
\times \left[ \left. \left(\frac{dN}{dt}\right)_{\mathrm{sat}}\right\vert_{\omega_0 \to 2\omega_0} \right]}, \\[4mm]
\displaystyle{
\frac{d}{dt} N_{\eta_1} \simeq 15 \left( \frac{M}{ \omega_0}\right)^2 \left(\frac{g \phi_0}{\omega_0}\right)^2  
\times \left(\frac{dN}{dt}\right)_{\mathrm{sat}}}, \\[4mm]
\displaystyle{
\frac{d}{dt} N_{\eta_2} \simeq 14 \left(\frac{g \phi_0}{\omega_0}\right)^2 \times \left(\frac{dN}{dt}\right)_{\mathrm{sat}}}, 
\end{cases}
\qquad \mathrm{for}\ \displaystyle{ \frac{M}{\omega_0} \ll 1,\  \frac{g\phi_0}{\omega_0} \ll 1,\  R\omega_0 =\pi}. 
\label{low mass result small phi}
\end{equation}

\begin{figure}[htbp]
\begin{center}
\setlength{\tabcolsep}{3pt}
\begin{tabular}{l l}
\resizebox{43mm}{!}{\includegraphics{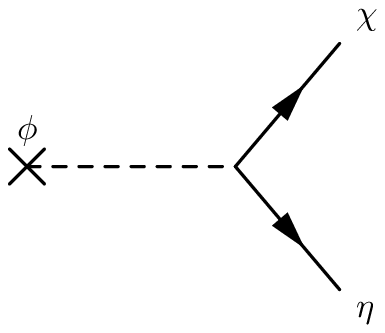}} &
\resizebox{43mm}{!}{\includegraphics{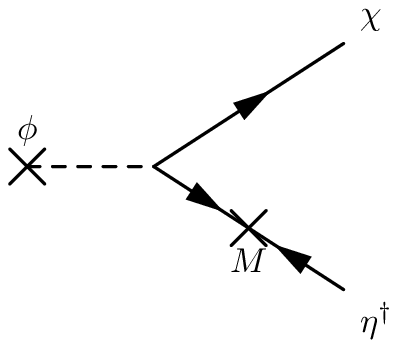}} \\
\end{tabular}
\end{center}
\caption{Diagrams for $\phi \to \chi \eta$.}
\label{diagram1}
\end{figure}

We can understand the above behavior of the production rates in the following way. 
Using the Feynman rules in Fig.~\ref{diag3}, 
we have the effective interaction, which is a good approximation only in the limit of $g \phi_0 /\omega_0 \ll 1$, as 
\begin{equation}
 \frac{1}{2} (g \phi_0)_\chi^{\mathrm{eff}} \chi \chi \simeq 
\begin{cases}
\displaystyle{
\left( \frac{g^2\phi^{*2}_0}{2M} \right) \chi \chi},  
\qquad & M/\omega_0 \gg 1, \\[4mm]
\displaystyle{
\left( \frac{g^2\phi^{*2}_0 M}{2 \omega_0^2}\right)  \chi \chi}, 
\qquad & M/\omega_0 \ll 1. 
\end{cases}
\end{equation}
Then, substituting this effective coupling $(g \phi_0)_\chi^{\mathrm{eff}}$ into Eq.~(\ref{rate small phi}) gives 
$dN_\chi/dt \simeq 12 (g^2\phi_0^2 M/\omega_0^3)^2 (R \omega_0-1.9) 
\times (dN/dt)_\mathrm{sat}$ for $M/\omega_0 \ll 1$. 
This is consistent with our numerical result of $dN_\chi/dt$ in Eq.~(\ref{low mass result small phi}) 
if we replace $(dN/dt)_\mathrm{sat}$ with $\left[ (dN/dt)_{\mathrm{sat}}\vert_{\omega_0 \to 2\omega_0} \right]$. 
We also consider the behavior of $dN_{\eta_i}/dt$. 
In the limit of $M/\omega_0 \ll 1$, Eq.~(\ref{eta trans}) becomes 
\begin{equation}
\begin{pmatrix}
\eta_1 \\
\eta_2 
\end{pmatrix}
\sim
\begin{pmatrix}
-\frac{M}{2\omega_0} & 1 \\
-1 & \frac{M}{2\omega_0}
\end{pmatrix}
\begin{pmatrix}
\eta \\
\eta^\dagger 
\end{pmatrix}, 
\label{eta trans lim}
\end{equation}
where we naively take $k_\eta$ as the typical energy $\omega_0$. 
From this and Fig.~\ref{diagram1}, we have the effective interactions as 
\begin{equation}
\begin{cases}
\left( (g\phi_0)_{\eta_1}^\mathrm{eff} \chi \eta_1 \right) \sim 
\left[ -M/2\omega_0 \left( (g \phi_0)_\eta^\mathrm{eff} \chi \eta \right)  + 
\left( (g \phi_0)^\mathrm{eff}_{\eta^\dagger} \chi \eta^\dagger \right) \right] 
&\to (g\phi_0)_{\eta_1}^\mathrm{eff} \sim M g\phi_0 /\omega_0, \\[2mm]
\left( (g\phi_0)_{\eta_2}^\mathrm{eff} \chi \eta_2 \right) \sim \left( -(g\phi_0)_\eta^\mathrm{eff} \chi \eta \right)   
&\to (g\phi_0)_{\eta_2}^\mathrm{eff} \sim g\phi_0, 
\end{cases}
\end{equation}
for $M/\omega_0 \ll 1$ and $g\phi_0/\omega_0 \ll 1$. 
Then, substituting these effective couplings $(g\phi_0)^\mathrm{eff}$ into Eq.~(\ref{rate small phi}), 
we obtain $dN_{\eta_1}/dt \simeq 12 (g \phi_0 M/\omega_0^2)^2 (R \omega_0 -1.9) \times (dN/dt)_\mathrm{sat}$ 
and $dN_{\eta_2}/dt \simeq 12 (g \phi_0 /\omega_0)^2 (R \omega_0-1.9) \times (dN/dt)_\mathrm{sat}$. 
These are consistent with our numerical result of $dN_{\eta_i}/dt$ in Eq.~(\ref{low mass result small phi}). 
Our results indicate that the effective theory gives us correct results even for the decay rate of the $Q$ ball. 

\section{\label{sec6}applications}
In the GMSB, $Q$ balls can decay only into gravitinos and hadrons if the next LSP mass is larger than $\omega_0$. 
We apply our results in the previous sections to gravitino and quark production from $Q$ balls in the GMSB 
and derive the branching ratio of the $Q$ ball decay into them. 
If the gravitino mass is small compared to the typical interaction energy, 
the longitudinal components of the gravitino behaves like the massless goldstino. 
Since the typical interaction energy is $\omega_0 =O(\mathrm{GeV}) \gg m_{3/2}$ in the GMSB, 
the effective interaction of Eq.~(\ref{goldstino}) is a good approximation to calculate the gravitino production rate. 
In this case, we have 
\begin{equation}
\begin{split}
\frac{\phi_0 \omega_0}{\langle F \rangle} 
&\simeq c \frac{m_s M_m}{g \langle F \rangle}, \\ 
& \simeq c \frac{g}{(4 \pi)^2} \frac{\langle F_s \rangle}{\langle F \rangle}. 
\end{split}
\end{equation}
Here we have used Eq.~(\ref{GMSB property}) and $m_s M_m \simeq \alpha \langle F_s \rangle /4 \pi$, 
where $\alpha=g^2/4\pi$ is the fine structure constant, and 
$\langle F_s \rangle$ is a vacuum expectation value 
for the $F$ component of a gauge-singlet chiral multiplet in the messenger sector. 
The SUSY breaking scale $ \langle F \rangle$ needs not be the same as the scale $\langle F_s \rangle$, 
i.e. $\langle F_s \rangle \leq \langle F \rangle$. 
From Eqs.~(\ref{HNO const}) and (\ref{result GMSB diff}), the gravitino production rate is calculated as 
\begin{equation}
\begin{split}
\frac{d}{dt}N_{\mathrm{gravitino}} &\simeq 0.9 \left( 4.8\log \frac{m_s}{\omega_0} +7.4 \right)^2  
\left( \frac{g}{(4\pi)^2} \right)^2 \left( \frac{\langle F_s \rangle}{\langle F \rangle} \right)^2
\times \left( \frac{dN}{dt} \right)_{\mathrm{sat}}, \\ 
&\simeq 0.7 \alpha \left( \frac{\langle F_s \rangle}{\langle F \rangle} \right)^2 
\times \left( \frac{dN}{dt} \right)_{\mathrm{sat}}. 
\end{split}
\end{equation}
Here and hereafter we take $m_s / \omega_0=10^3$. 
There is also massive gluino. The Lagrangian density is written as 
\begin{equation}
\mathcal{L}_{gluino}=\mathcal{L}_{kinetic}-\sqrt{2} g {\phi}^* \lambda q -\frac{1}{2} 
M_g \lambda \lambda + h.c.
\end{equation}
where $\lambda$ and $q$ are gluino and quark, respectively, and $M_g$ is the gluino mass. 
Typically, $M_g = O(\mathrm{TeV})$ and $\omega_0 = O(\mathrm{GeV})$, so we can use Eq.~(\ref{resultmassive}). 
Using $m_s, M_m  \gg \omega_0$ and $M_g \sim m_s$ in Eq.~(\ref{GMSB property}), 
we can see that 
$\left( \omega_0/M_g\right) \left( \sqrt{2} g \phi_0/\omega_0 \right)^2 \sim  m_s^2 M_m^2  / (M_g \omega_0^3 ) \gg 1$. 
Thus, the quark production rate is calculated from Eqs.~(\ref{GMSB rate})~and~(\ref{resultmassive}) as 
\begin{equation}
\frac{d}{dt}N_{\mathrm{quark}} 
\simeq 1.1 \times \left[ \left. \left( \frac{dN}{dt}\right)_{\mathrm{sat}}\right\vert_{\omega_0 \to 2\omega_0} \right]. 
\end{equation}
We conclude that the main decay channel is the decay into quarks and is saturated, and 
the branching ratio of the decay into the gravitino is calculated as 
\begin{equation}
\begin{split}
B_{3/2} 
&\simeq 
0.1 \left( 4.8\log \frac{m_s}{\omega_0} +7.4 \right)^2 \left( \frac{g}{(4\pi)^2}\right)^2 
\left( \frac{\langle F_s \rangle}{\langle F \rangle} \right)^2, \\ 
&\simeq 0.08 \alpha \left( \frac{\langle F_s \rangle}{\langle F \rangle} \right)^2. 
\label{B3/2result}
\end{split}
\end{equation}
This branching ratio can be rewritten as 
\begin{equation}
\begin{split}
B_{3/2} &\simeq 
0.1 \left( 4.8\log \frac{m_s}{\omega_0} +7.4 \right)^2 
\left( \frac{m_s M_m}{\sqrt{3} g m_{3/2} M_{\mathrm{P}}} \right)^2, \\ 
&\simeq \frac{4}{\alpha} \left( \frac{m_s M_m}{m_{3/2} M_{\mathrm{P}}}\right)^2, 
\end{split}
\end{equation}
where we use $m_s M_m  \simeq g^2 \langle F_s \rangle /(4 \pi)^2$ and 
$\langle F \rangle = \sqrt{3} m_{3/2} M_{\mathrm{P}}$ 
($M_{\mathrm{P}}=2.4 \times 10^{18}\mathrm{GeV}$: the reduced Planck mass). 

We compare the above branching ratio with the one estimated in Ref.~\cite{KK2011}. 
The quark production rate was estimated from the effective coupling 
$g'_{\mathrm{eff}} \simeq g^2 \phi_0/\sqrt{2\pi M_g \omega_0}$ 
for the process squark $+$ squark $\to$ quark $+$ quark via gluino exchange. 
Since we have $g'_{\mathrm{eff}} \phi_0 /\omega_0 \gg 1$, the quark production rate is saturated. 
The gravitino production rate was estimated from the effective coupling 
$g_{\mathrm{eff}} \simeq  \omega_0^2/\sqrt{2}  \langle F \rangle$ 
because the elementary process squark $\to$ quark $+$ gravitino has the decay rate 
$\Gamma = m_\phi^5 / (16\pi \langle F \rangle^2)$. 
Thus, the branching ratio of the decay into the gravitino is estimated from Eq.~(\ref{Cohen un sat}) as~\cite{KK2011} 
\begin{equation}
B'_{3/2} \simeq 3 \pi \frac {g_{\mathrm{eff}} \phi_0}{\omega_0} 
\simeq 0.7 \sqrt{\alpha} \left( \frac{\langle F_s \rangle}{\langle F \rangle} \right) 
\simeq \frac{5}{\sqrt{\alpha}} \frac{m_s M_m}{m_{3/2} M_{\mathrm{P}}}. 
\end{equation}
However, Eq.~(\ref{Cohen un sat}) can not be applied to the case of $R\omega_0 \sim 1$, 
which is the case in the GMSB, and we should use Eqs.~(\ref{rate small phi})~and~(\ref{GMSB rate}). 
In addition, we should use $(dN/dt)_{\mathrm{sat}} \vert_{\omega_0 \to 2\omega_0}$ for the quark production rate 
because the produced quark energy is in the interval $(0, 2\omega_0)$. 
If we take into account these considerations and use $g_{\mathrm{eff}}$, 
the branching ratio is estimated as 
\begin{equation}
B''_{3/2} \simeq \frac{3}{2 \pi^2} 12 \left( \frac {g_{\mathrm{eff}} \phi_0}{\omega_0} \right)^2 
(R_Q \omega_0-1.9) \times \frac{1}{8} 
\simeq 0.14 \frac{ c^2 g^2}{ (4\pi)^4} \left( \frac{\langle F_s \rangle}{\langle F \rangle} \right)^2 
\simeq 0.12 \alpha \left( \frac{\langle F_s \rangle}{\langle F \rangle} \right)^2, 
\label{B3/2estimate2}
\end{equation}
where Eqs.~(\ref{GMSBR})~and~(\ref{GMSB property}) are used in the second equality. 
This result has the same parameter dependences with our numerical result of Eq.~(\ref{B3/2result}), 
and the numerical factor is also correct within the order of one. 
This shows that the naive use of the effective coupling is a good approximation 
even for the decay rate of the $Q$ ball into gravitinos. 

\section{\label{sec7}conclusions}
We have calculated the fermion production rates from the step-function type, 
the gauge-mediation type and the gravity-mediation type of $Q$ ball in the Yukawa theory. 
In the limit of $g\phi_0/\omega_0 \ll 1$, we have found that the decay rates can be obtained from 
the step-function type of $Q$ ball with the correction factors coming from the relations between 
the total charge $Q$ and the radius $R \simeq 1/ \omega_0$. 
On the other hand, the decay rates in the limit of $g\phi_0/\omega_0 \gg 1$ are saturated 
and proportional to the square of the effective $Q$ ball radius $R'$, 
where $R' $ is determined by $g\phi( R')/\omega_0\sim 1$. 

We have also calculated the goldstino production rates 
from the gauge-mediation type and the gravity-mediation type of $Q$ ball 
using the low energy interaction with the supercurrent. 
Our results can be explained by the production rate through the Yukawa interaction 
where the Yukawa coupling $g$ is replaced by the effective coupling 
$g_{\mathrm{eff}} \simeq \omega_0^2 / \sqrt{2} \langle F \rangle$. 
This effective coupling comes from the fact that the elementary process squark $\to$ quark $+$ gravitino 
has the decay rate $\Gamma = m_\phi^5 / (16\pi \langle F \rangle^2)$ 
or we can naively estimate derivatives in the interaction term as 
$\partial / \partial r \sim 1/R$ and $\partial / \partial t \sim \omega_0$. 

We have also calculated the $Q$ ball decay rates in the Yukawa theory with a massive fermion. 
Our results are consistent with the effective theory 
once we make the replacement $\omega_0 \to 2\omega_0$ in the saturated rate of the massless fermion, 
since the produced fermion energy is in the interval $(0, 2\omega_0)$. 
Especially, when $\omega_0 < M$, we can integrate out the heavy particle and 
use the effective coupling $(g \phi_0)^{\mathrm{eff}} \simeq g^2\phi_0^2/2M$ in the Yukawa theory 
with the replacement $\omega_0 \to 2\omega_0$ in the saturated rate. 

In the GMSB model, the branching ratio of the decay into the gravitino has been calculated as 
\begin{equation}
\begin{split}
B_{3/2} 
&\simeq 
0.1  \left( 4.8\log \frac{m_s}{\omega_0} +7.4 \right)^2 \left( \frac{g}{(4\pi)^2}\right)^2 
\left( \frac{\langle F_s \rangle}{\langle F \rangle} \right)^2, \\ 
&\simeq 0.08 \alpha \left( \frac{\langle F_s \rangle}{\langle F \rangle} \right)^2,  \\ 
&\simeq \frac{4}{\alpha} \left( \frac{m_s M_m}{m_{3/2} M_{\mathrm{P}}}\right)^2, 
\end{split}
\label{conclusion}
\end{equation}
for $m_s / \omega_0=10^3$. 
This branching ratio is much less than the one estimated in Ref.~\cite{KK2011}. 
The main reason is that 
Ref.~\cite{KK2011} used the production rate for the limit of $R\omega_0 \to \infty$ which is not 
valid in the GMSB. 
Another reason is that we should take into account the $Q$ ball configuration which is different from a step function. 
Therefore, the gravitino dark matter from the $Q$ balls in the GMSB 
should be reconsider using the correct decay rates obtained in the present paper, 
which will be presented elsewhere~\cite{KKY}. 

\section*{Acknowledgment}

We thank Shinta Kasuya for useful discussions. 
This work is supported by Grant-in-Aid for Scientific research from
the Ministry of Education, Science, Sports, and Culture (MEXT), Japan,
No.\ 14102004 (M.K.), No.\ 21111006 (M.K.) and also 
by World Premier International Research Center
Initiative (WPI Initiative), MEXT, Japan.




\begin{thebibliography}{90}

\bibitem{AD}
  I.~Affleck and M.~Dine,
  Nucl.\ Phys.\  B {\bf 249}, 361 (1985).

\bibitem{DRT} 
  M.~Dine, L.~Randall and S.~D.~Thomas,
  Nucl.\ Phys.\ B {\bf 458}, 291 (1996).

\bibitem{KuSh}
  A.~Kusenko and M.~E.~Shaposhnikov,
  Phys.\ Lett.\  B {\bf 418}, 46 (1998).
  
\bibitem{EnMc}
  K.~Enqvist and J.~McDonald,
  Phys.\ Lett.\  B {\bf 425}, 309 (1998); 
%
  Nucl.\ Phys.\  B {\bf 538}, 321 (1999).
  
\bibitem{KK1}
  S.~Kasuya and M.~Kawasaki,
  Phys.\ Rev.\  D {\bf 61}, 041301(R) (2000).

\bibitem{KK2}
  S.~Kasuya and M.~Kawasaki,
  Phys.\ Rev.\  D {\bf 62}, 023512 (2000).

\bibitem{KK3}
  S.~Kasuya and M.~Kawasaki,
  Phys.\ Rev.\  D {\bf 64}, 123515 (2001).

\bibitem{KLS1} 
  A.~Kusenko, L.~Loveridge and M.~Shaposhnikov,
  Phys.\ Rev.\ D {\bf 72}, 025015 (2005)

\bibitem{KLS2} 
  A.~Kusenko, L.~C.~Loveridge and M.~Shaposhnikov,
  JCAP {\bf 0508}, 011 (2005)
  
\bibitem{ShKu}
  I.~M.~Shoemaker and A.~Kusenko,
  Phys.\ Rev.\  D {\bf 80}, 075021 (2009).
  
\bibitem{DoMc} 
  F.~Doddato and J.~McDonald,
  JCAP {\bf 1106}, 008 (2011).
  
\bibitem{KK2011}
S.~Kasuya and M.~Kawasaki,
  Phys.\ Rev.\ D {\bf 84}, 123528 (2011).

\bibitem{evap}
  A.~G.~Cohen, S.~R.~Coleman, H.~Georgi and A.~Manohar,
  Nucl.\ Phys.\  B {\bf 272}, 301 (1986).

\bibitem{HNO} 
  J.~Hisano, M.~M.~Nojiri and N.~Okada,
  Phys.\ Rev.\ D {\bf 64}, 023511 (2001).

\bibitem{Coleman}
S.~Coleman,
Nucl.\ Phys.\ {\bf B262} (1985) 263.

\bibitem{Qsusy}
A.~Kusenko,
Phys.\ Lett.\ {\bf B405} (1997) 108.

\bibitem{log2}
  A.~de Gouv\^ea, T.~Moroi and H.~Murayama,
  Phys.\ Rev.\  D {\bf 56}, 1281 (1997).

\bibitem{MV} 
  T.~Multamaki and I.~Vilja,
  Nucl.\ Phys.\ B {\bf 574}, 130 (2000).

\bibitem{KKY} 
  S.~Kasuya, M.~Kawasaki and M.~Yamada, 
  in preparation. 

\end{thebibliography}
\end{document}